\documentclass[a4paper,pre,twocolumn]{revtex4-1}
\usepackage[dvips]{graphicx}
\usepackage{amsmath}
\usepackage{psfrag,color}
\usepackage{amssymb}
\usepackage{natbib}
\usepackage{ulem}
\usepackage{multirow}
\usepackage{booktabs}
\usepackage{mathrsfs} 

\newcommand{\upd}{\mathrm{d}}
\newcommand{\drho}{\Delta\rho}

\begin{document}
\title{The compression of a heavy floating elastic film}

\author{Etienne Jambon-Puillet$^a$, Dominic Vella$^{a,b}$ and Suzie Proti\`ere$^a$}
\email[]{protiere@ida.upmc.fr}
\affiliation{ $^a$Institut Jean le Rond d'Alembert, UPMC Paris 6, CNRS UMR 7190, 4 Pl Jussieu, 75005, Paris,  France\\$^b$Mathematical Institute, Andrew Wiles Building, University of Oxford, Woodstock Rd, Oxford, OX2 6GG, UK}
\date{\today}
\begin{abstract}
We study the effect of film density on the uniaxial compression of thin elastic films at a liquid--fluid interface. Using a combination of experiments and theory, we show that dense films first wrinkle and then fold as the compression is increased, similarly to what has been reported when the film density is neglected. However, we highlight the changes in the shape of the fold induced by the film's own weight and extend the model of Diamant and Witten [\emph{Phys.~Rev.~Lett.}, 2011, \textbf{107}, 164302] to understand these changes. In particular, we suggest that it is the weight of the film that breaks the up-down symmetry apparent from previous models, but elusive experimentally. We then compress the film beyond the point of self-contact and observe a new behaviour dependent on the film density: the single fold that forms after wrinkling transitions into a closed loop after self-contact, encapsulating a cylindrical droplet of the upper fluid. The encapsulated drop either causes the loop to bend upward or to sink deeper as compression is increased, depending on the relative buoyancy of the drop-film combination. We propose a model to qualitatively explain this behaviour. Finally, we discuss the relevance of the different buckling modes predicted in previous theoretical studies and highlight the important role of surface tension in the shape of the fold that is observed from the side --- an aspect that is usually neglected in theoretical analyses.
\end{abstract}

\maketitle

\section{Introduction}

The formation of wrinkles and their localization into folds is observed when a layered material is progressively compressed. As such, they occur in a variety of systems at different scales in Nature from the folding of geological strata \cite{Pollard2005} to  the morphogenesis of biological tissues such as the cerebral cortex\cite{VanEssen1997, Toro2005, Tallinen2016} or fingerprints \cite{Kucken2004} as well as in the growth of biofilms \cite{Trejo2013}. Just as nature uses these mechanical instabilities to generate intricate patterns, wrinkling instabilities are also finding a number of technological applications  from flexible electronic devices \cite{Rogers2010, Kim2010a} to controlled patterned surfaces \cite{Bowden1998, Schweikart2009} that have improved light harvesting efficiency \cite{Kim2012b} or surface hydrophobicity \cite{Lin2015}. Because of this ubiquity and the renewed interest in taking control of elastic instability, the deformation of layered elastic materials has been the object of renewed interest recently, with a number of fundamental studies in the past decade \cite{Cerda2003, Bo2012, Reis2009, Brau2011, Kim2011}.

The simplest example of elastic instability is the classic Euler buckling \cite{Levien2008,Landau} in which a compressed beam buckles over its entire length. To use elastic instability for pattern formation usually requires the selection of a length scale other than the system size, however. Perhaps the simplest example of such a system is provided by an elastic film (of bending stiffness per unit width $B$) floating on a liquid of density $\rho$ and subject to an axial compression. This system quickly forms wrinkles with a wavelength $\lambda \sim (B/\rho g)^{1/4}$ --- a result that expresses the compromise between the bending rigidity of the film (which prefers large-amplitude wrinkles) and the weight of the liquid ($g$ being the gravitational acceleration), which prefers many small-amplitude wrinkles\cite{Pocivavsek2008, Holmes2010, King2012, Pineirua2013}. This wrinkling instability can also be observed in other types of floating materials such as thin layers of nanoparticles\cite{Leahy2010}, surfactants \cite{Lu2002, Boatwright2010} or particle rafts \cite{Vella2004}. However, at some point most of these wrinkles begin to shrink with a very small number growing to form much larger `folds'; although other options (such as delamination\cite{Wagner2011}) exist, folds are more generically observed and so we focus on them in this paper.

The nature of the wrinkle-to-fold transition in this system has been somewhat controversial: it was initially suggested (on the basis of experiments) that folds form at a critical compression. Several numerical and analytical works have studied the wrinkle-to-fold transition in idealized situations (in which  the elastic film is infinitely long and weightless \cite{Diamant2011, Diamant2013, Brau2013, Rivetti2013}) and have shown that in these scenarios the transition is not sharp: i.e.~folds emerge continuously from the wrinkled state. Recently, a more realistic model (incorporating the finite length of the film) demonstrated that this transition emerges at a finite compression making a second-order transition \cite{Rivetti2014, Oshri2015}. Despite the amount of work on the wrinkle-to-fold transition, several questions remain unanswered. For example, existing models suggest that downwards and upwards folds should be equally favourable, while experiments report exclusively downward folds. Moreover, what happens beyond the transition between these two states has not been described experimentally but only via numerical simulations\cite{Demery2014} with the weight of the film discussed in scaling terms. In the first part of this paper we investigate the effect of the film weight on the wrinkle-to-fold transition while in the second part we focus on the evolution of the fold upon further compression. We observe that the fold becomes a closed loop after self-contact, encapsulating a volume of the upper fluid. As the applied compression is increased, we observe that the fold either sinks deeper without limit or eventually bends upward, depending on the relative buoyancy of the drop-film combination. We predict the regime diagram for this behaviour theoretically.

\section{Experimental section}
\textbf{Elastic film production:} Thin elastic films are produced by spincoating vinylpolysiloxane VPS (elite double 32, \texttt{Zhermack}) of base density $\rho_{VPS}=1.20\mathrm{~g\,cm^{-3}}$. To produce films of various densities, we add iron powder (97\%, -325 mesh, \texttt{Sigma Aldricht}) $\rho_{Fe}=7.87 \mathrm{~g\,cm^{-3}}$ to the non cross-linked polymer. This allows us to work with films of densities $\rho_s=1.2-2.6\mathrm{~g\,cm^{-3}}$ (see table, SI). The spin-coated polymer/iron mixture viscosity changes with the iron powder concentration so that at a fixed rotation speed the film thickness, $t$, obtained varies. We adjust the rotation speed to narrow the thickness range ($50\mathrm{~\mu m}<t<120 \mathrm{~\mu m}$). Two sets of film lengths $L_0$ and width $W$ are produced: $L_0=90\mathrm{~mm}$, $W=60\mathrm{~mm}$ and $L_0=75\mathrm{~mm}$, $W=50\mathrm{~mm}$.

\noindent\textbf{Compression experiment:} 

\begin{figure}
\centering
  \includegraphics[width=8.3cm]{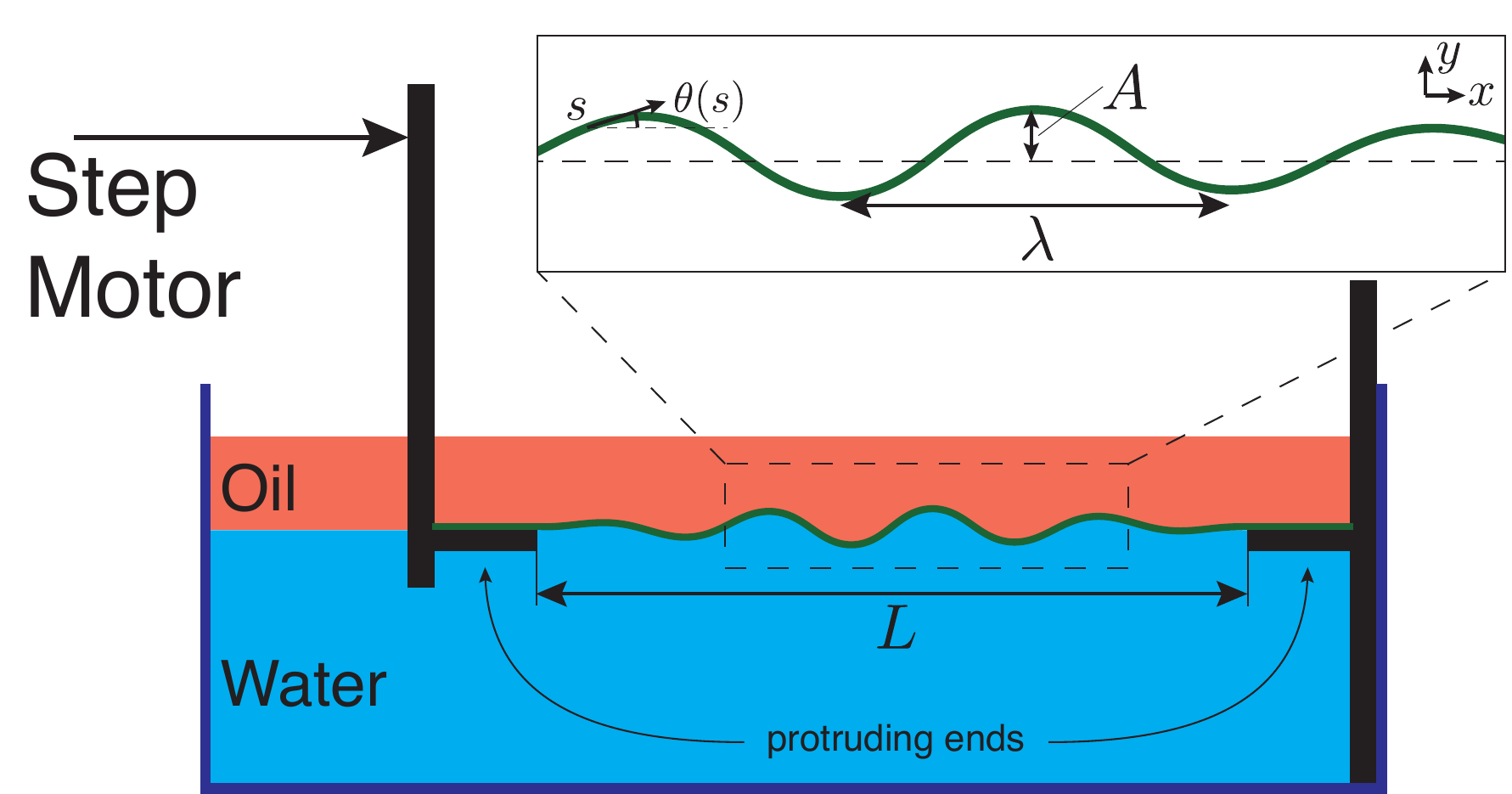}
  \caption{Schematic of the experimental setup defining the length $L$ between the boundaries, the wrinkle wavelength $\lambda$ and the wrinkle amplitude $A$ as well as the Cartesian and intrinsic coordinates used in the model: ($x$,$y$) and ($s$,$\theta$).}
  \label{fgr:manip}
\end{figure}

The experiments are conducted in a glass tank ($12 \times 11 \times 6.5\mathrm{~cm^3}$) with two parallel polymethylmethacrylate (PMMA) plates with horizontal protruding ends (Fig.~\ref{fgr:manip}). The tank is filled with tap water to a level higher than the protruding ends. The elastic film is then carefully placed at the air/water interface between the plates. The water level is then lowered (using a syringe) until the edges of the film comes into contact with the protruding ends of the PMMA plates. The film naturally adheres strongly to the PMMA achieving a clamped boundary condition at the protruding ends. The water level is adjusted with the syringe so that the film is completely flat. A light mineral oil ($\rho_o=0.836\mathrm{~g\,cm^{-3}}$, \texttt{Sigma Aldrich}) is then poured slowly on top of the film (so that no oil invades the lower surface of the film, in contact with water). One of the PMMA plates is mounted on two perpendicular manual translation stages for alignment in the ($y$,$z$) direction and connected to a stepper motor (\texttt{Thorlabs}) of micrometer precision. The compression is quasi-static: the stepper motor displaces the plate in small increments at a constant speed. The motor stops for $5\mathrm{s}$ between each step allowing the system to relax to its equilibrium shape. The elastic film is imaged from the side and/or the top with two \texttt{Nikon} D800-E cameras mounted with macro objectives ($105$ mm). The images are then analysed using either \texttt{ImageJ} or \texttt{MATLAB}.

\noindent\textbf{Film analysis:}

We use two methods to determine the elastic films thicknesses. We weigh the films on a milligram scale: knowing the density, length and width we deduce an average thickness. We also illuminate the films with a tilted laser line. The laser deflection being a function of the height and tilting angle, height profiles are then extracted using the appropriate calibration. Both measurement techniques give similar results. In addition, the laser line allows us to check that our films are uniform: thickness variations are below the measurement standard deviation ($\pm\:10\mathrm{~\mu m}$) except at the edges of the film where a small ridge ($20\mathrm{~\mu m}$ thicker) develops during the spin coating/curing. The Young's modulus $E$ and Poisson ratio $\nu$ are determined by a tensile test on a dogbone shape using a \texttt{Shimadzu} tensile machine. For the pure VPS we find $E=1.0\mathrm{~MPa}$, $\nu=0.5$ while for the VPS/iron particle mixture the Young's modulus increases up  to $E=2.9\mathrm{~MPa}$ for a dogbone of density $\rho_s=2.5\mathrm{~g\,cm^{-3}}$. We checked that the film Young's modulus is the same as that obtained from the test samples by measuring the deflection of the film under its own weight. By letting a small portion of the film protrude out of a clamp we were able to vary the length easily and fit the bending stiffness $B=Et^3/\big[12(1-\nu^2)\bigr]$. The values found through this procedure are in good agreement with the previously measured $E$ and $t$. Finally the measured $E$ and $t$ are compatible with the direct measurement of the wrinkle wavelength $\lambda$ (see table, SI).


\section{Results}

\begin{figure}
  \includegraphics[width=8.3cm]{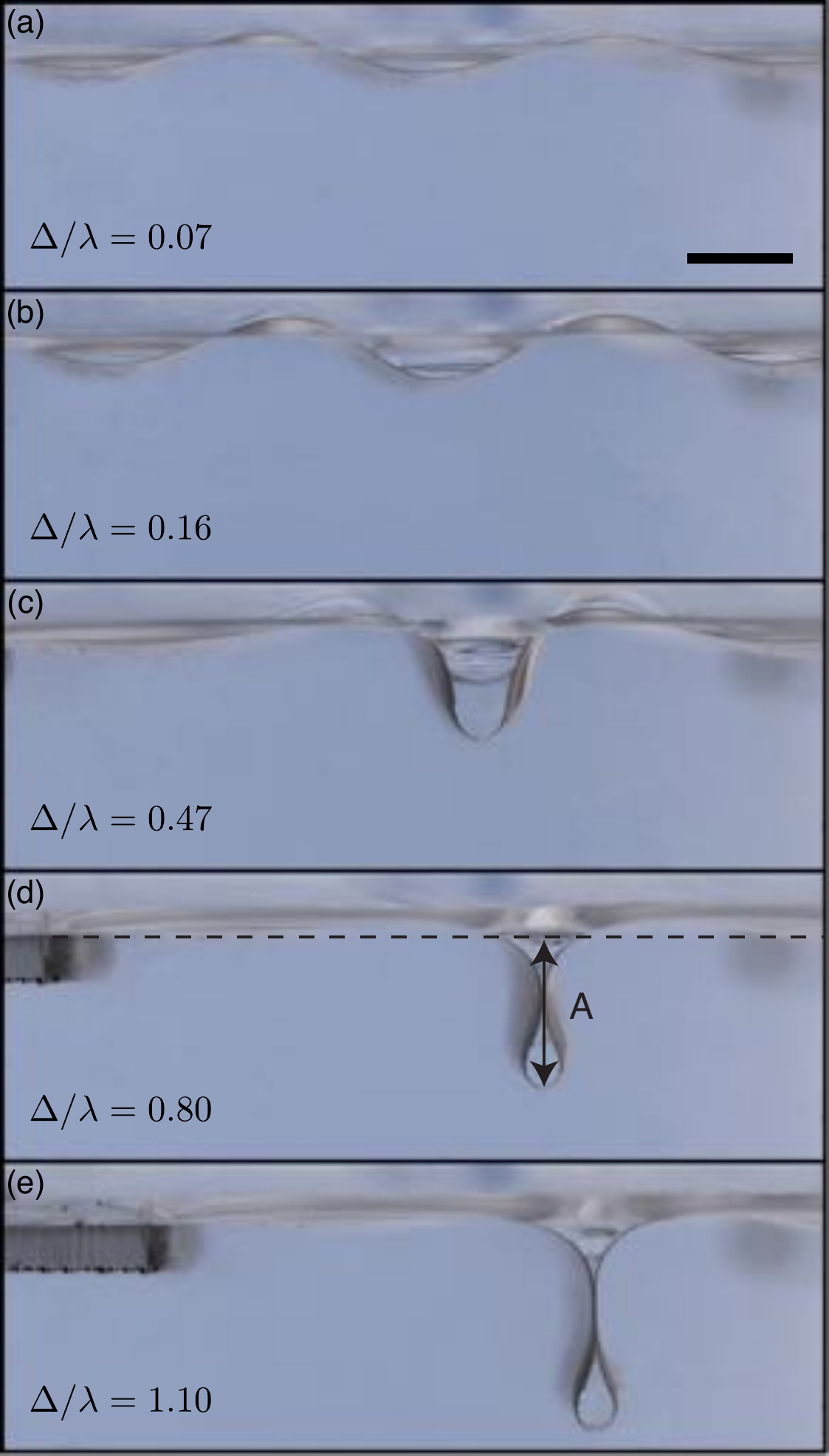}
  \caption{Side images of an elastic film of density $\rho_s=1.8\:g.cm^{-3}$ at an oil/water interface. Compression $\Delta$ increases from \textbf{(a)} to \textbf{(e)}, scale bar 5mm. Colours have been inverted to enhance contrast. \textbf{(a)-(b)} The film displays a quasi-periodic out of plane displacement of amplitude $A$ and wavelength $\lambda$, referred to as the wrinkled state. \textbf{(c)} The deformation localizes in a single fold. \textbf{(d)} The fold reaches self-contact. \textbf{(e)} A column of oil is encapsulated in the fold which grows deeper towards the bottom of the tank as compression is increased.
  }
  \label{FigEvolution}
\end{figure}

We compress an elastic film floating at an interface between two fluids and study its behaviour as we increase the imposed displacement. At zero compression, $L=L_0$, the film lies flat at the interface. As soon as we start compressing the film ($\Delta=L_0-L>0$), it buckles out of plane with a characteristic wavelength $\lambda$ that develops along the film length (Fig.~\ref{FigEvolution} (a)). This is the wrinkled state. As $\Delta$ increases, the wrinkle amplitude  $A$ initially grows uniformly (Fig.~\ref{FigEvolution} (b)) until, suddenly, only one of the wrinkles continues to grow while the other wrinkles   progressively vanish. The excess  length of the vanished wrinkles is absorbed by the remaining  structure, the fold (Fig.~\ref{FigEvolution}(c)). We therefore observe a clear transition between the ``wrinkled state", in which the deformation is distributed throughout the film, and a ``fold state", in which all the deformation is localized in a narrow region of high curvature, \textit{i.e.}: the fold. As the compression $\Delta$ is increased beyond the wrinkle-to-fold transition, the fold continues to grow in amplitude and its curvature increases. Finally, the two opposing edges of the fold come  into contact  (self-contact). At this point a horizontal column of the upper fluid is encapsulated in this ``teardrop" shape (Fig.~\ref{FigEvolution} (d)). With further increases in compression this teardrop is forced down into the subphase and does not noticeably change shape; instead the length over which the film is in self-contact increases (Fig.~\ref{FigEvolution} (e)). In this paper, we  study the role of the density of the elastic film in the wrinkle-to-fold transition and its impact on the evolution of the fold as the uniaxial compression is increased to the point of self-contact and beyond.

\subsection{Wrinkle-to-fold transition and fold before self-contact}

\begin{figure}
  \centering
  \includegraphics[width=8.3cm]{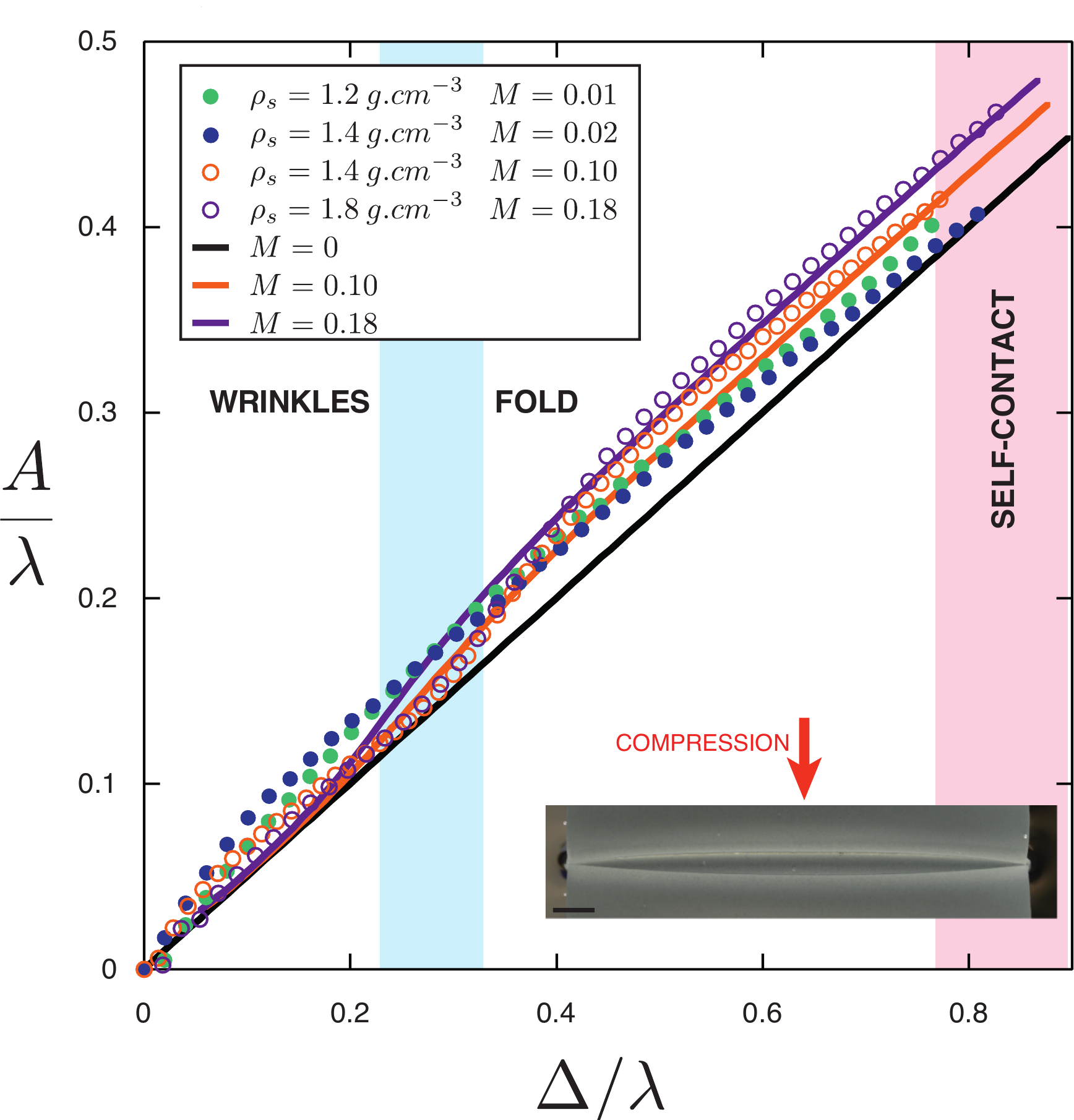}
  \caption{Dimensionless wrinkle/fold amplitude as a function of the dimensionless compression up to self-contact. (\protect\includegraphics[height=2mm]{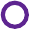}, \protect\includegraphics[height=2mm]{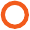}) represent oil/water experiments, (\protect\includegraphics[height=2mm]{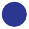}, \protect\includegraphics[height=2mm]{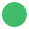}) air/water experiments. Three film densities are presented (with different lengths/widths/thicknesses), giving rise to the four dimensionless weights presented in the legend. The black curve corresponds to the symmetric solution \cite{Diamant2011} ($M=0$), the orange and purple curves correspond to the symmetric numerical solution of equation \eqref{nlsystM} for $M=0.10$ and $M=0.18$. \textit{Inset}: Top view of the compressed film. The fold reaches self-contact at the edges of the film but is still open in the centre (see SI). Scale bar: 5mm.}
  \label{fgr:Amppresc}
\end{figure}

 In order for the fold to nucleate near the centre of the elastic film, at $L_0/2$, we ensure that the clamped edges are aligned very carefully. (The fold has to appear at least one wavelength $\lambda$ away from the clamped edges to avoid any boundary effects.) The results were compared to a fold generated by applying a small pressure at the film centre before starting to compress it. For three different densities $\rho_s=1.2,\:1.4,\:1.8\mathrm{~g\,cm^{-3}}$ and two different film lengths $L_0$, we used  images taken from the side to measure the fold amplitude $A$ during compression up to self-contact. The data are then normalized by the wavelength $\lambda$ measured experimentally (Fig.~\ref{fgr:Amppresc}). For small compressions, $\Delta / \lambda \lesssim 0.3$ we observe the wrinkled state with no influence of the film weight. In this wrinkled regime, Pocivavsek \textit{et al.}~\cite{Pocivavsek2008} found that the evolution of the wrinkle amplitude exhibits a weak dependence  on the film length, which we are not able to observe in our experiments. Moreover, Rivetti and Neukirch \cite{Rivetti2014} also predict that the buckling modes (symmetric/antisymmetric/non symmetric) of the film can evolve during the compression for different film lengths. We do observe two different buckling routes in our experiments and they are responsible for the two different trends in the amplitude that we observe in Fig.~\ref{fgr:Amppresc}: the experiments at an air--water interface remain symmetric throughout the compression while the experiments at an oil--water interface start antisymmetric but switch continuously to a symmetric mode. Since the antisymmetric mode has a smaller amplitude (for a given compression) than the symmetric one, when the fold switches to the symmetric mode its amplitude increases rapidly. This results in an inflection point at $\Delta/\lambda\sim 0.3$ for experiments at an oil--water interface. However, these details  are very sensitive to the experimental conditions (particularly to the clamped boundary conditions) and no quantitative comparison could be made with the predictions of Rivetti and Neukirch \cite{Rivetti2014} (see SI). For larger compressions,  $\Delta / \lambda > 0.4$, the behaviour does not depend on film length: the deformation always localizes in a downward symmetric fold with an amplitude that grows linearly with compression, as described previously \cite{Pocivavsek2008, Diamant2011}.

However, as we vary the film densities we find that the amplitude of the fold increases slightly with the film mass (Fig.~\ref{fgr:Amppresc}). To explain this dependence we first turn to the model developed by Diamant and Witten \cite{Diamant2011} for the nonlinear deflections of a floating elastic film. In the limit of an incompressible, weightless film of infinite length, this model has  an analytical solution\cite{Diamant2011}; here we shall have to develop a numerical solution but we first outline the derivation of the governing equation for a light film, so that the appropriate modifications can be made for the film weight. We introduce the intrinsic coordinates ($s$,$\theta$), where $\theta$ is the angle between the film and the horizontal axis and $s$ is the arc-length;  the film centreline is then parametrized in terms of arc-length, $[x(s),y(s)]$. The energy $U$ of a film of bending stiffness $B$ and width $W$ contains contributions from bending $U_b=\tfrac{1}{2}BW\int_{-\infty}^{\infty} \left(\partial_s \theta\right)^2~\upd s$ and from the gravitational potential energy of the underlying fluid substrate $U_s$. In our case the gravitational energy of the upper and lower fluids  $U_{l_{up}}$ and $U_{l_{low}}$, respectively are given by: $U_s=U_{l_{up}}+U_{l_{low}}=\tfrac{1}{2}\drho gW\int_{-\infty}^{\infty} y(s)^2\cos{\theta}~\upd s$ where the relevant density is the density difference between the two fluids $\drho=\rho_{low}-\rho_{up}$. The energy is to be minimized subject to the constraint of an imposed compression $\Delta=\int_{-\infty}^{\infty}(1-\cos\theta)~\upd s$. To facilitate the calculation, lengths are  non-dimensionalized by $\ell_{eh}=\bigl[B/(\drho g)\bigr]^{1/4}=\lambda/(2\pi)$ and energies by $WB/\ell_{eh}$. The minimization itself is discussed in detail by Diamant and Witten\cite{Diamant2011} and yields a single equation for the intrinsic angle $\theta(s)$ with the boundary conditions $\theta (\pm \infty) = \partial_s \theta (\pm \infty) = y (\pm \infty) =0 $:
\begin{equation}
\partial_s^4 \theta + \bigl[\tfrac{3}{2}\left(\partial_s \theta\right)^2+P\bigr] \partial_s^2 \theta + \sin \theta=0
\label{Witteneq}
\end{equation}
Here $P$ is the Lagrange multiplier associated with the compression constraint, and  corresponds physically to the dimensionless horizontal force applied on the elastic film by the compression. Equation \eqref{Witteneq} has a family of analytical solutions \cite{Diamant2011, Diamant2013, Rivetti2013}:
\begin{equation}
\begin{split}
\theta & (s)=4\arctan \bigg( \frac{\kappa \sin\bigl[k(s+\phi)\bigr]}{k \cosh(\kappa s)} \bigg)\\
k=\tfrac{1}{2} & \sqrt{2+P} , \quad \kappa=\tfrac{1}{2}\sqrt{2-P}, \quad P=2-\frac{\Delta^2}{16}.
\end{split}
\label{Wittensol}
\end{equation} This family of solutions is parametrized by $\phi$, with $0 \le k\phi \le \pi$; the value of $k\phi$ selects the symmetry of the profile\cite{Rivetti2013}, with $k\phi=0$ corresponding to an even fold and $k\phi=\pi/2$ corresponding to an odd fold\cite{Diamant2011}.

For the experiments presented in this paper, the weight of the film is not negligible; we must therefore supplement  the bending and substrate energy with the  gravitational energy of the film itself, $U_g$. If $\rho_s$ is the density of the film and $\rho_{w}$ the density of the lower liquid (water) then the dimensionless gravitational energy of the film is
\[U_g = M \int_{-\infty}^{\infty} y(s)~\upd s,\]
where
\begin{equation}
M=\frac{(\rho_s-\rho_w)t}{\drho \ell_{eh}}
\label{eqn:Mdefn}
\end{equation} is a dimensionless number that compares the weight of the film to the restoring force provided by buoyancy over the horizontal length $\ell_{eh}$. Applying the energy minimization procedure described by \cite{Diamant2011} to the total energy $U=U_s+U_b+U_g$  we obtain the system of equations (see SI~for details):
\begin{eqnarray}
&\partial_s^4 \theta + \left[\tfrac{3}{2}\left(\partial_s \theta\right)^2+P-My\right] \partial_s^2 \theta +(1-2M\partial_s \theta)\sin \theta=0 \nonumber \\
&\partial_s x=\cos \theta, \quad \partial_s y=\sin \theta \label{nlsystM} 
\end{eqnarray}

using the same clamped boundary conditions as for the light film. We note that, as expected, the weightless equation  \eqref{Witteneq} is recovered by setting $M=0$.

For $M\neq0$, we are not able to find an analytical solution of the system of equations \eqref{nlsystM}; instead we obtain numerical solutions by using the \texttt{MATLAB} routine \texttt{bvp4c}, which uses a relaxation technique with the analytical solution \eqref{Wittensol} as an initial guess. We also use a simple  continuation algorithm to find the solutions as parameters are varied.

Our numerical solutions suggest that the antisymmetric solution no longer exists once $M\neq0$ (we are unable to find numerical solutions with this symmetry).  Even if an antisymmetric solution of equation \eqref{nlsystM} does exist, with $M>0$ (respectively $M<0$), the downward (respectively upward) symmetric solution has a lower energy due to the contribution of the gravitational term $U_g$. (In the limit $M=0$,  all of the solutions \eqref{Wittensol} are known to have the same energy \cite{Diamant2011, Diamant2013, Brau2013, Rivetti2013}.) The weight of the film therefore lifts the degeneracy that was inferred from the analytical solution \eqref{Wittensol} but this has not been observed experimentally\cite{Pocivavsek2008}; previously this breaking of symmetry was attributed to surface tension, but not quantified \cite{Pocivavsek2008}.

Experimentally, we do, in fact, observe antisymmetric configurations for sufficiently low compressions ($\Delta / \lambda \lesssim 0.25$); however, as the compression increases, the film always evolves to a downward symmetric configuration (see SI). This is in accordance with the theoretical prediction that  the energy difference between the two states increases  with $\Delta / \lambda$.

The evolution of the amplitude with increasing compression within the symmetric solution, as determined from the numerical solution of \eqref{nlsystM} for $M=0,\:0.10,\:0.18$, is shown in Fig.~\ref{fgr:Amppresc}. For $\Delta / \lambda \lesssim 0.2$, the system remains in the wrinkled state and there is very little influence of the film weight on the amplitude--compression curve. However, with $\Delta / \lambda > 0.25$, the fold amplitude increases with the film weight $M$, holding  compression, $\Delta$, fixed; furthermore, we recover the linear evolution of the fold amplitude with $\Delta$. As $M$ varies, the experimentally determined $(\Delta,A)$ curves are shifted as predicted by our theory (Fig.~\ref{fgr:Amppresc}). The  observed experimental fold is fully localized and downward symmetric for $\Delta / \lambda > 0.4$. Its amplitude can thus be quantitatively predicted by the numerical solution of \eqref{nlsystM} in this regime. 

In fig.~\ref{fgr:Amppresc}, the last point of each curve corresponds to the point at which the downward symmetric fold first self-contacts, forming  a loop, \textit{i.e.}~ the teardrop shape. Experimentally, we find that this point is reached for $\Delta_{sc}/\lambda \approx 0.8$ whereas using equation \eqref{nlsystM}, this transition is found at a slightly higher value ($\Delta_{sc}/\lambda \approx 0.89$ for symmetric folds). The reason for this discrepancy is not immediately clear.
Some insight into this shift is obtained by examining a top view of the fold (inset of Fig.~\ref{fgr:Amppresc}): its shape is not the same at the edge of the film and at its centre . Capillary forces pull the elastic film on its sides closing the fold tighter at its edges than at its centre. Surface tension slightly distorts the fold's shape at the sheet's edges, which explains the small discrepancy between the theoretical and experimental $\Delta_{sc}$. However the fold's amplitude remains constant even when subjected to surface tension effects (see SI~for details).

\subsection{Evolution of the fold}

\begin{figure}
\centering
  \includegraphics[width=8.3cm]{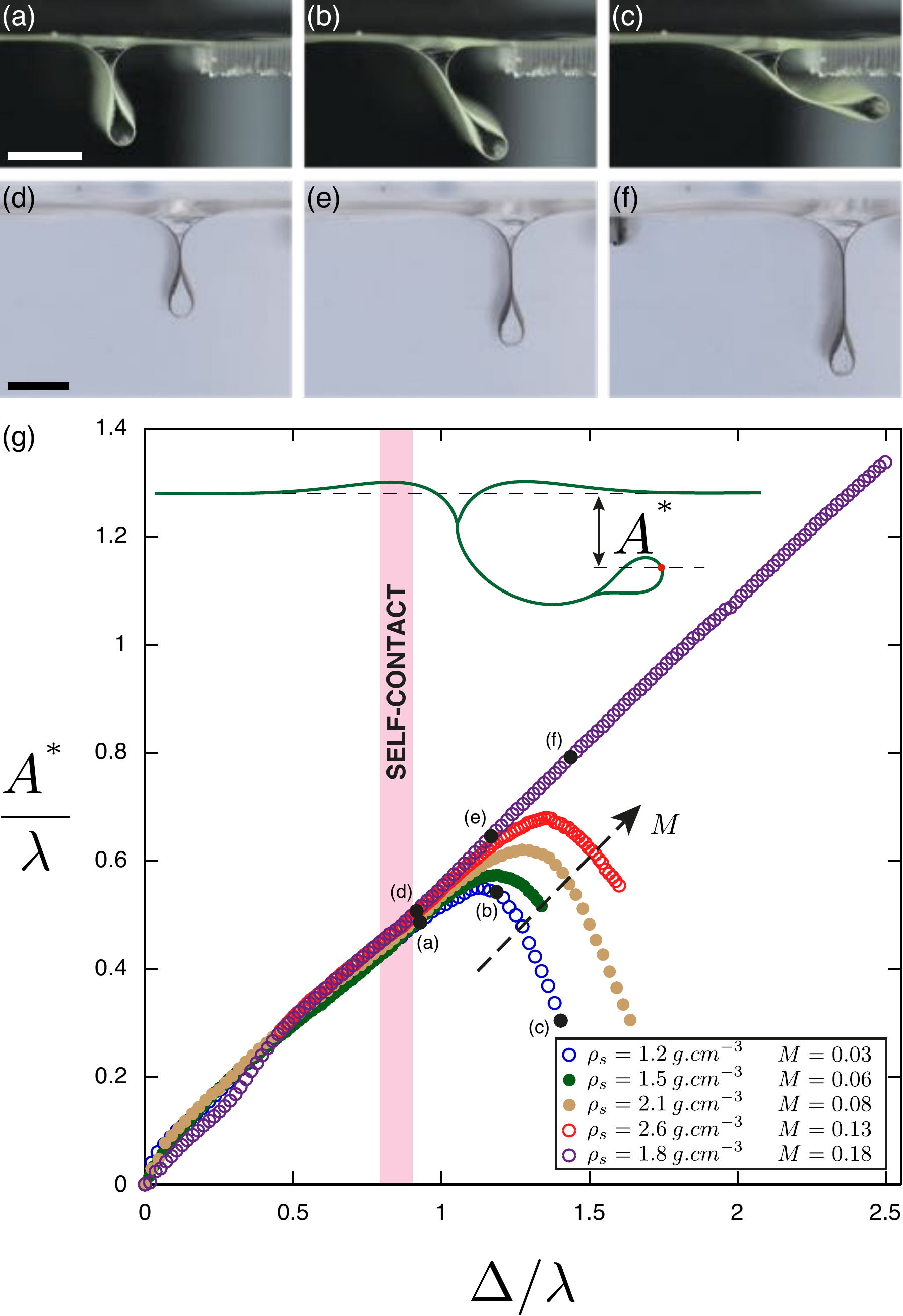}
  \caption{Images from the side of two folds as compression is increased. The two sets of pictures have the same dimensionless compressions, from left to right $\Delta/\lambda \approx \: 0.93,\: 1.19,\: 1.44$. Scalebars: $5\mathrm{~mm}$. \textbf{(a)-(c)} film with a low mass $M=0.03$, $\rho_s=1.20\mathrm{~g\,cm^{-3}}$. \textbf{(d)-(e)} film with a high mass  $M=0.18$, $\rho_s=1.8\mathrm{~g\,cm^{-3}}$. Colours have been inverted to enhance contrast. \textbf{(g)} \textit{Inset}: Schematic presenting the post buckling fold amplitude $A^*$ (fold centre in red). Dimensionless wrinkle/fold amplitude $A^*$ as a function of the dimensionless compression after self-contact. (\protect\includegraphics[height=2mm]{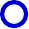}, \protect\includegraphics[height=2mm]{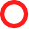}, \protect\includegraphics[height=2mm]{op_purp_circ.pdf}) represents oil/water experiments, (\protect\includegraphics[height=2mm]{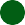}, \protect\includegraphics[height=2mm]{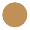}) air/water experiments. \protect\includegraphics[height=2mm]{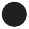} show the data corresponding to the pictures \textbf{(a)-(f)}. }
  \label{fgr:amppostsc}
\end{figure}

The analytical solutions \eqref{Wittensol} and the numerical solutions of \eqref{nlsystM} cease to be relevant beyond self-contact ($\Delta>\Delta_{sc}$) since, within the limitations of our model, the numerically generated profiles interpenetrate. Experimentally, we observe two different behaviours depending on the dimensionless mass $M$, defined in \eqref{eqn:Mdefn}, as shown in Fig.~\ref{fgr:amppostsc} (a)-(f): heavy films ($M\gtrsim0.14$) retain a symmetric configuration and encapsulate a column of the upper fluid in a teardrop shape fold (Fig.~\ref{fgr:amppostsc} (d)-(e)) that grows deeper as the compression increases (Fig.~\ref{fgr:amppostsc} (f)). Lighter films ($M\lesssim0.14$) start symmetric with the same teardrop shape but, as the compression increases, the loop starts to tilt and grows back up towards the interface (Fig.~\ref{fgr:amppostsc} (a)-(b)) where it finally reaches an obstacle such as the clamped edges or itself (Fig.~\ref{fgr:amppostsc} (c)). To quantify these observations, we first  define the amplitude after self-contact, $A^*$, as the depth of the centre of the loop, see the inset of Fig.~\ref{fgr:amppostsc} (g). This quantity is clearly defined beyond self-contact, while before self-contact,  $A^*\equiv A$. Fig.~\ref{fgr:amppostsc} (g) shows the experimentally measured amplitude as a function of the compression before and after self-contact. We vary the film density and length and present our results using the dimensionless mass $M$ described previously. At the scale of Fig.~\ref{fgr:amppostsc} the variation of amplitude due to the mass $M$ is not visible before self-contact; immediately after self-contact the amplitude initially keeps growing linearly ($A^*/\lambda \approx 0.5\Delta/\lambda$). However, as the compression is increased beyond self-contact, the influence of $M$ becomes more apparent, at a critical compression $\Delta_b$  the fold begins to tilt back up towards the interface with increasing compression \textit{i.e.}: $A^*$ decreases. (We emphasize that this tilting occurs quasi-statically.) Fig.~\ref{fgr:amppostsc} (g) shows that as $M$ increases, the transition from a straight to a tilted fold occurs at larger $\Delta_b$. Finally, when the elastic film is sufficiently dense ($M=0.18$ in Fig.~\ref{fgr:amppostsc} (g)), the fold  never bends back up and sinks indefinitely into the liquid subphase.

\begin{figure}
\centering
  \includegraphics[width=0.45\textwidth]{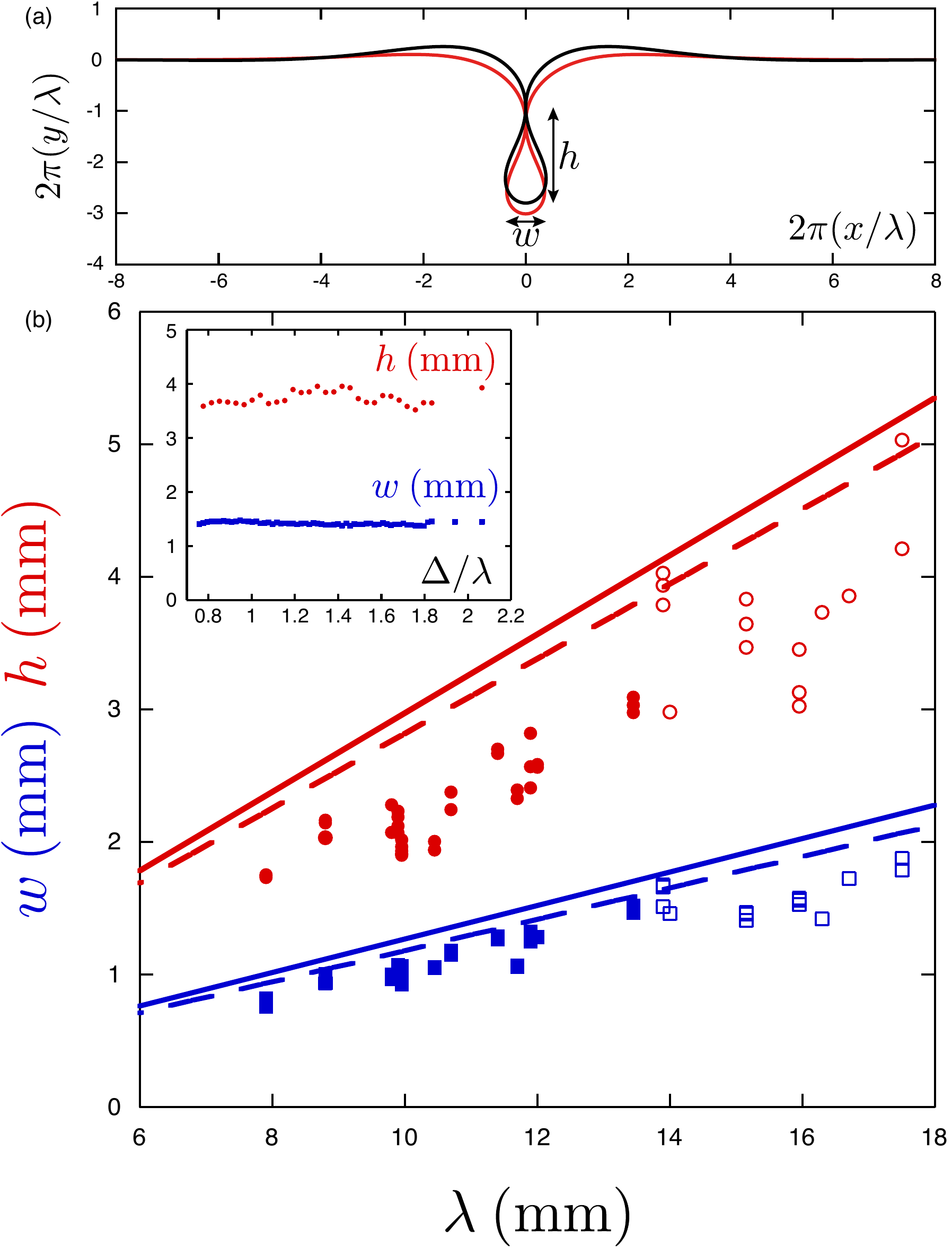}
  \caption{\textbf{(a)} Comparison of symmetric profiles at self-contact given by the analytical solution \eqref{Wittensol} for $M=0$ (black) and the shape for $M=0.18$, determined from the numerical solution of equation \eqref{nlsystM} (red). The arrows define the height $h$ and the width $w$ of the teardrop. \textbf{(b)} The teardrop height $h$ (red circles) and width $w$ (blue squares) as a function of the wavelength. Closed symbols represent air/water experiments, open symbols oil/water experiments; here, solid lines correspond to the analytical prediction, based on equation \eqref{Wittensol}, while dashed lines correspond to the numerical solution of equation \eqref{nlsystM} for $M=0.18$. All other experimental parameters are found to be not relevant and vary across the data. \textit{Inset}: Experimental teardrop height and width in one experiment ($M=0.14$, $\rho_s=1.8\mathrm{~g\,cm^{-3}}$) as a function of the compression.}
  \label{fgr:formebouclelarge}
\end{figure}

To understand the tilting behaviour of the loop, we first consider its properties at the point of self-contact. The key quantities of interest are the loop's  width $w$ and height $h$. Experiments show that these grow linearly with $\lambda$ (see Fig.~\ref{fgr:formebouclelarge} (b)) but do not vary with compression beyond self-contact  (see inset of Fig.~\ref{fgr:formebouclelarge} (b)). We believe that this lack of evolution of the teardrop shape is due mainly to the constant volume of the upper liquid that is trapped in the loop after the point of self-contact. The teardrop shape observed corresponds to that predicted by the solution of equations \eqref{Wittensol} and \eqref{nlsystM} \emph{at} self-contact (Fig.~\ref{fgr:formebouclelarge} (a)). However, the values of $w$ and $h$ are slightly overestimated by the model since we can only measure the shape at the film edges, where surface tension effects come into play;  at the centre of the film the loop is wider (see inset Fig.~\ref{fgr:Amppresc}). The numerical solution of equation \eqref{nlsystM} predicts that the loop formed at self-contact shrinks as $M$ increases. We were not able to observe this experimentally, and conclude that  this effect must be smaller than the role of surface tension at the film edges. 

We compare our results to the numerical work by Demery \textit{et al.} \cite{Demery2014} where they study a compressed film  after self-contact; however, they do not incorporate the weight of the film itself. They find two possible configurations for the shape of the film in this case: a symmetric one (which we also observe experimentally) and an antisymmetric one (which we have never observed). In the weightless case the antisymmetric configuration has a lower energy --- its energy saturates with increasing $\Delta$, while the energy of the symmetric downward fold \emph{increases} linearly with $\Delta$. In our case the added weight of the film reduces the energy of the symmetric configuration by $U_g\sim-M \Delta^2$, explaining why this configuration is ultimately energetically favourable. 

To describe the evolution of the fold when $\Delta>\Delta_{sc}$ in more detail,  we need a new mathematical model. We continue to neglect surface tension  so that the film behaviour is invariant along its width and the problem can be treated as two-dimensional. Our model of the elastic film after self-contact is illustrated schematically in Fig.~\ref{fgr:heavyelexpdiagphase}. We break the system in two parts: a heavy beam corresponding to the part of the film in self-contact and a force $F$ acting at its tip (which is the teardrop shape encapsulating buoyant fluid). Here
\begin{equation}
F=\drho g \mathscr{A}-(\rho_s-\rho_{w}) g t \mathscr{L} ,
\label{fgr:force}
\end{equation}
$\mathscr{A}$ and  $\mathscr{L}$ are respectively the half-area and half the perimeter of  the upper fluid encapsulated in the teardrop (see schematic in the inset of Fig.~\ref{fgr:heavyelexpdiagphase}). Considering these half quantities means that we are considering each half of the two that are in contact separately, and assuming they do not exert force on one another. While this reduction suggests that the problem is equivalent to that of a beam subject to a constant load at one end, we emphasize that the self-weight of the beam is important and so instead we must consider a `heavy hanging column'\cite{Wang1981}, subject to a constant force pushing at the tip (since we have already found experimentally that the teardrop size does not evolve with further compression). We assume a clamped boundary condition at the top for simplicity (schematic Fig.~\ref{fgr:heavyelexpdiagphase}). At each compression step the portion in self-contact $L^*$ grows, increasing the length of the effective beam.

We introduce again  intrinsic coordinates ($s$,$\theta$), with $s$ the arc-length but now $\theta$ denotes the angle between the heavy beam and the vertical axis. The $(x, y)$ coordinates are thus rotated clockwise by $90$ degrees compared to the model presented in the previous section. In this system, the equation for the heavy hanging column is given by \cite{Wang1981, Wang1986}(see SI~for details):
 \[ B \partial_s^2 \theta = -[F-(\rho_s-\rho_{w})gts] \sin \theta\]
 $B$ is the beam bending modulus and $t$ its thickness. $\rho_s$ is the beam density and $\rho_{w}$ the lower liquid (water) density. The boundary conditions are that the end at  $s=0$ is free while that at  $s=L^*$ is clamped:
 \[ \partial_s \theta(s=0)=0, \quad \theta(s=L^*)=0\]
 The system is made dimensionless by dividing $s$ by $L^*$:
 \begin{equation}
\begin{split}
\partial_s^2 \theta + \left[\widetilde{F}-\bigg(\frac{L^*}{\ell_g}\bigg)^3 s\right]\sin \theta =0 \\
\end{split}
\label{heavyeleq}
\end{equation}
\[\partial_s \theta(s=0)=0, \quad \theta(s=1)=0.\]
Here $\widetilde{F}=F{L^{*}}^2/B$ is the dimensionless force due to the buoyancy of the teardrop and $\ell_g=(B/[(\rho_s-\rho_{w})gt])^{1/3}$ is the elasto-gravitational length, which compares elastic and gravitational forces; loosely speaking the elasto-gravitational length is the length above which the film will buckle under its own weight. The elasto-gravitational length is also closely related to the parameter $M$, which is defined in \eqref{eqn:Mdefn} and may be written $M=(\ell_{eh}/\ell_g)^3$.

For a fixed applied force $F$, the beam will buckle once the length reaches a threshold value. This threshold can be determined by linearizing \eqref{heavyeleq} and solving the resulting problem analytically  \cite{Wang1981} (see SI). This analysis yields the critical force $\widetilde{F}_c$ needed for the beam to buckle as a function of $(L^*/\ell_g)^3$, \textit{i.e.}~$\widetilde{F}_c=\frac{F_c{L^*}^2}{B}=\operatorname{f}\big((L^*/\ell_g)^3\big)$. In our experiments, the control parameter is the beam length $L^*$, which increases with increasing compression; the force is constant, since it originates from the buoyancy of the teardrop, which is fixed during the compression. We therefore write:
\begin{equation}
\frac{F\ell_g^2}{B}=\Big(\frac{\ell_g}{L^*_c}\Big)^2\operatorname{f}\bigg(\frac{L^*_c}{\ell_g}\bigg)
\label{heavyelthreshold2}
\end{equation}
where the function $\operatorname{f}(x)$ emerges from a solvability condition, see SI. For a given teardrop size, and hence buoyancy force, equation \eqref{heavyelthreshold2} may be solved to give the critical length at which the beam begins to bend.

\begin{figure}
\centering
  \includegraphics[width=8.3cm]{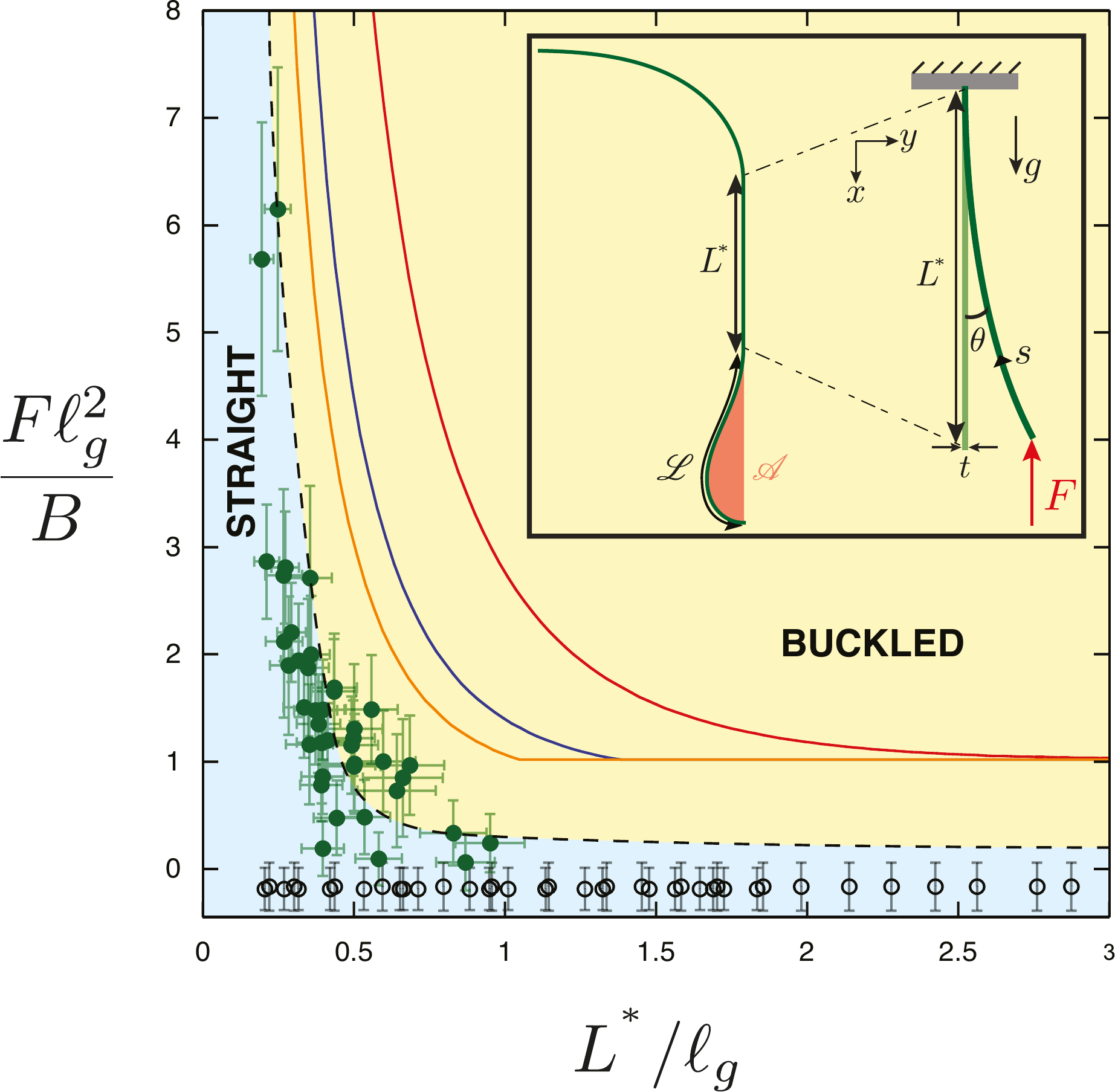}
  \caption{Phase diagram of the post self-contact fold buckling. Green filled circles give the critical length $L^*$ before buckling in each experiment. Black open circles represent experiments in which the fold never buckles. The black dashed curve and background colours are guides for the eye to distinguish experimental phases. The red solid curve is the analytical result of the heavy hanging column (equation \eqref{heavyelthreshold2}), the blue (respectively orange) solid curve is the numerical result of the heavy hanging column theory with an initial angle of 10$^{\circ}$ (respectively 20$^{\circ}$). \textit{Inset}: Schematic of a perfectly symmetric fold post self-contact. The fold is split in two parts: the portion in self-contact is treated as a heavy hanging column of length $L^*$ and thickness $t$ and the teardrop is modeled as a constant point force (resulting from the buoyancy of the teardrop). The upper part is not taken into account here and a clamped boundary condition is assumed at the top of the heavy hanging column. The intrinsic coordinates ($s$,$\theta$) are drawn.}
  \label{fgr:heavyelexpdiagphase}
\end{figure}

The computed prediction for the length at which the fold should start to bend upwards is presented in Fig.~\ref{fgr:heavyelexpdiagphase}. Here, the buoyancy force of the teardrop is calculated using equation \eqref{fgr:force} with $\mathscr{A}$, $\mathscr{L}$ and $L^*$ measured from images taken from the side (and are hence subject to the inaccuracies discussed earlier) and $B$ is calculated based on the measurement of the film wrinkling wavelength ($B=\drho g \bigl[\lambda/(2 \pi)\bigr]^4$). This calculation, via equation \eqref{heavyelthreshold2}, captures qualitatively the experimentally observed fold behaviour. In particular, there is a critical buoyancy force below which the fold never buckles, regardless of the length of the film. 

The experimental data points shown in Fig.~\ref{fgr:heavyelexpdiagphase} lie well below the prediction of equation \eqref{heavyelthreshold2}, although the trend is similar. We believe that in most experiments this discrepancy occurs because the fold is not perfectly aligned with the vertical axis when it reaches self-contact (Fig.~\ref{fgr:amppostsc}(a)). The average angle $\alpha$ between the fold and the vertical axis is $\left|\alpha\right|=10^{\circ}$. To account for this angle, we may include this effect by changing the boundary condition at the top of the beam: $\theta(s=1)=\alpha$. We now solve the problem numerically (using equation \eqref{heavyeleq}) to determine the critical length  $L^*_c/\ell_g$ for buckling (see SI~for details). The results are plotted in Fig.~\ref{fgr:heavyelexpdiagphase} (blue curve: $\alpha=10^{\circ}$, orange curve: $\alpha=20^{\circ}$) and are closer to the experimental values. An additional source of discrepancy between the experimental results and the numerical model may be, once again,  that we underestimate the size of the teardrop, since we measure its shape at the edge of the film (where it is affected by surface tension) and not at its centre. It may also be due to how we oversimplify the boundary condition at the top of the beam. Here, we have assumed that  a clamped boundary condition is appropriate but we have observed that the top of the fold can slide and rotate slightly as the compression is varied.

\section{Conclusion}
We have studied experimentally the behaviour of thin elastic films under uniaxial compression at a liquid-fluid interface. As previously observed, when confinement increases the film undergoes a transition from a uniform wrinkled state to a localized configuration in which a single fold is observed. We showed that the film weight affects the wrinkle-to-fold transition, breaking the up-down symmetry of the light film problem considered previously. More importantly, however, the film weight plays a large role in the evolution of the fold beyond the transition: it may bend back towards the fluid interface or grow deeper as compression is increased. A simple model predicts the two possible behaviours in a phase diagram.

The experimental results presented here allow us to confirm some  previous theoretical assumptions but also highlight the role of the film weight and surface tension in the formation and evolution of a fold. The teardrop shape observed after self-contact seems similar to the so-called ``self-encapsulation" of an elastic rod which is observed in a solely elastic situation when a rod covering a fixed span is loaded at the middle with a transverse force such that two points of the rod come into contact with each other\cite{Bosi2015}. Finally, this study could give insight into the elastic properties of compressed particle-laden interfaces, which exhibit a wrinkling instability similar to that described for elastic films in this Letter. For example, our results show that such films may encapsulate a teardrop of the upper (lower) fluid if the dimensionless parameter $M>0$ ($M<0$); this may provide a novel route for the creation of cylindrical, particle-coated droplets \cite{Subramaniam2005} by compressing a particle-laden interface. In such settings, the granular aspect of the interface allows the interface's mechanical properties to be tuned by probing new parameters such as the particle size, density or packing fraction \cite{Petit2016} as well as the solid-liquid contact angles. However, the granular aspect also affects the stress state within the layer at the onset of wrinkling \cite{Cicuta2009} meaning that the transition from wrinkles to localized folds upon further compression remains to be carefully characterized in such systems.

\section{Acknowledgement}
We are grateful to S\'ebastien Neukirch, Vincent D\'emery and Claude Perdigou for useful discussions as well as Aur\'elie Fargette for her help with the preparation of thin films and Julien Chopin for his expertise with the tensile testing. The research leading to these results has received funding from the European Research Council under the European Union's Horizon 2020 Programme/ ERC Grant Agreement no. 637334 (DV).


\begin{thebibliography}{41}%
\makeatletter
\providecommand \@ifxundefined [1]{%
 \@ifx{#1\undefined}
}%
\providecommand \@ifnum [1]{%
 \ifnum #1\expandafter \@firstoftwo
 \else \expandafter \@secondoftwo
 \fi
}%
\providecommand \@ifx [1]{%
 \ifx #1\expandafter \@firstoftwo
 \else \expandafter \@secondoftwo
 \fi
}%
\providecommand \natexlab [1]{#1}%
\providecommand \enquote  [1]{``#1''}%
\providecommand \bibnamefont  [1]{#1}%
\providecommand \bibfnamefont [1]{#1}%
\providecommand \citenamefont [1]{#1}%
\providecommand \href@noop [0]{\@secondoftwo}%
\providecommand \href [0]{\begingroup \@sanitize@url \@href}%
\providecommand \@href[1]{\@@startlink{#1}\@@href}%
\providecommand \@@href[1]{\endgroup#1\@@endlink}%
\providecommand \@sanitize@url [0]{\catcode `\\12\catcode `\$12\catcode
  `\&12\catcode `\#12\catcode `\^12\catcode `\_12\catcode `\%12\relax}%
\providecommand \@@startlink[1]{}%
\providecommand \@@endlink[0]{}%
\providecommand \url  [0]{\begingroup\@sanitize@url \@url }%
\providecommand \@url [1]{\endgroup\@href {#1}{\urlprefix }}%
\providecommand \urlprefix  [0]{URL }%
\providecommand \Eprint [0]{\href }%
\providecommand \doibase [0]{http://dx.doi.org/}%
\providecommand \selectlanguage [0]{\@gobble}%
\providecommand \bibinfo  [0]{\@secondoftwo}%
\providecommand \bibfield  [0]{\@secondoftwo}%
\providecommand \translation [1]{[#1]}%
\providecommand \BibitemOpen [0]{}%
\providecommand \bibitemStop [0]{}%
\providecommand \bibitemNoStop [0]{.\EOS\space}%
\providecommand \EOS [0]{\spacefactor3000\relax}%
\providecommand \BibitemShut  [1]{\csname bibitem#1\endcsname}%
\let\auto@bib@innerbib\@empty
\bibitem [{\citenamefont {Pollard}\ and\ \citenamefont
  {Fletcher}(2005)}]{Pollard2005}%
  \BibitemOpen
  \bibfield  {author} {\bibinfo {author} {\bibfnamefont {D.~D.}\ \bibnamefont
  {Pollard}}\ and\ \bibinfo {author} {\bibfnamefont {R.~C.}\ \bibnamefont
  {Fletcher}},\ }\href@noop {} {\emph {\bibinfo {title} {Fundamentals of
  Structural Geology}}}\ (\bibinfo  {publisher} {Cambridge University Press},\
  \bibinfo {year} {2005})\BibitemShut {NoStop}%
\bibitem [{\citenamefont {Van~Essen}(1997)}]{VanEssen1997}%
  \BibitemOpen
  \bibfield  {author} {\bibinfo {author} {\bibfnamefont {D.~C.}\ \bibnamefont
  {Van~Essen}},\ }\href@noop {} {\bibfield  {journal} {\bibinfo  {journal}
  {Nature}\ }\textbf {\bibinfo {volume} {385}},\ \bibinfo {pages} {313}
  (\bibinfo {year} {1997})}\BibitemShut {NoStop}%
\bibitem [{\citenamefont {Toro}\ and\ \citenamefont {Burnod}(2005)}]{Toro2005}%
  \BibitemOpen
  \bibfield  {author} {\bibinfo {author} {\bibfnamefont {R.}~\bibnamefont
  {Toro}}\ and\ \bibinfo {author} {\bibfnamefont {Y.}~\bibnamefont {Burnod}},\
  }\href {\doibase 10.1093/cercor/bhi068} {\bibfield  {journal} {\bibinfo
  {journal} {Cerebral Cortex}\ }\textbf {\bibinfo {volume} {15}},\ \bibinfo
  {pages} {1900} (\bibinfo {year} {2005})}
  \BibitemShut {NoStop}%
\bibitem [{\citenamefont {Tallinen}\ \emph {et~al.}(2016)\citenamefont
  {Tallinen}, \citenamefont {Chung}, \citenamefont {Rousseau}, \citenamefont
  {Girard}, \citenamefont {Lefevre},\ and\ \citenamefont
  {Mahadevan}}]{Tallinen2016}%
  \BibitemOpen
  \bibfield  {author} {\bibinfo {author} {\bibfnamefont {T.}~\bibnamefont
  {Tallinen}}, \bibinfo {author} {\bibfnamefont {J.~Y.}\ \bibnamefont {Chung}},
  \bibinfo {author} {\bibfnamefont {F.}~\bibnamefont {Rousseau}}, \bibinfo
  {author} {\bibfnamefont {N.}~\bibnamefont {Girard}}, \bibinfo {author}
  {\bibfnamefont {J.}~\bibnamefont {Lefevre}}, \ and\ \bibinfo {author}
  {\bibfnamefont {L.}~\bibnamefont {Mahadevan}},\ }\href@noop {} {\bibfield
  {journal} {\bibinfo  {journal} {Nat. Phys.}\ }\textbf {\bibinfo {volume}
  {12}},\ \bibinfo {pages} {588} (\bibinfo {year} {2016})}\BibitemShut
  {NoStop}%
\bibitem [{\citenamefont {K\"ucken}\ and\ \citenamefont
  {Newell}(2004)}]{Kucken2004}%
  \BibitemOpen
  \bibfield  {author} {\bibinfo {author} {\bibfnamefont {M.}~\bibnamefont
  {K\"ucken}}\ and\ \bibinfo {author} {\bibfnamefont {A.~C.}\ \bibnamefont
  {Newell}},\ } {\bibfield
  {journal} {\bibinfo  {journal} {EPL (Europhysics Letters)}\ }\textbf
  {\bibinfo {volume} {68}},\ \bibinfo {pages} {141} (\bibinfo {year}
  {2004})}\BibitemShut {NoStop}%
\bibitem [{\citenamefont {Trejo}\ \emph {et~al.}(2013)\citenamefont {Trejo},
  \citenamefont {Douarche}, \citenamefont {Bailleux}, \citenamefont {Poulard},
  \citenamefont {Mariot}, \citenamefont {Regeard},\ and\ \citenamefont
  {Raspaud}}]{Trejo2013}%
  \BibitemOpen
  \bibfield  {author} {\bibinfo {author} {\bibfnamefont {M.}~\bibnamefont
  {Trejo}}, \bibinfo {author} {\bibfnamefont {C.}~\bibnamefont {Douarche}},
  \bibinfo {author} {\bibfnamefont {V.}~\bibnamefont {Bailleux}}, \bibinfo
  {author} {\bibfnamefont {C.}~\bibnamefont {Poulard}}, \bibinfo {author}
  {\bibfnamefont {S.}~\bibnamefont {Mariot}}, \bibinfo {author} {\bibfnamefont
  {C.}~\bibnamefont {Regeard}}, \ and\ \bibinfo {author} {\bibfnamefont
  {E.}~\bibnamefont {Raspaud}},\ }\href {\doibase 10.1073/pnas.1217178110}
  {\bibfield  {journal} {\bibinfo  {journal} {Proc. Natl. Acad. Sci. U. S. A.}\
  }\textbf {\bibinfo {volume} {110}},\ \bibinfo {pages} {2011} (\bibinfo {year}
  {2013})}\BibitemShut {NoStop}%
\bibitem [{\citenamefont {Rogers}\ \emph {et~al.}(2010)\citenamefont {Rogers},
  \citenamefont {Someya},\ and\ \citenamefont {Huang}}]{Rogers2010}%
  \BibitemOpen
  \bibfield  {author} {\bibinfo {author} {\bibfnamefont {J.~A.}\ \bibnamefont
  {Rogers}}, \bibinfo {author} {\bibfnamefont {T.}~\bibnamefont {Someya}}, \
  and\ \bibinfo {author} {\bibfnamefont {Y.}~\bibnamefont {Huang}},\ }\href
  {\doibase 10.1126/science.1182383} {\bibfield  {journal} {\bibinfo  {journal}
  {Science}\ }\textbf {\bibinfo {volume} {327}},\ \bibinfo {pages} {1603}
  (\bibinfo {year} {2010})}\BibitemShut
  {NoStop}%
\bibitem [{\citenamefont {Kim}\ \emph {et~al.}(2010)\citenamefont {Kim},
  \citenamefont {Kim}, \citenamefont {Xiao}, \citenamefont {Kim}, \citenamefont
  {Park}, \citenamefont {Panilaitis}, \citenamefont {Ghaffari}, \citenamefont
  {Yao}, \citenamefont {Li}, \citenamefont {Liu}, \citenamefont {Malyarchuk},
  \citenamefont {Kim}, \citenamefont {Le}, \citenamefont {Nuzzo}, \citenamefont
  {Kaplan}, \citenamefont {Omenetto}, \citenamefont {Huang}, \citenamefont
  {Kang},\ and\ \citenamefont {Rogers}}]{Kim2010a}%
  \BibitemOpen
  \bibfield  {author} {\bibinfo {author} {\bibfnamefont {R.-H.}\ \bibnamefont
  {Kim}}, \bibinfo {author} {\bibfnamefont {D.-H.}\ \bibnamefont {Kim}},
  \bibinfo {author} {\bibfnamefont {J.}~\bibnamefont {Xiao}}, \bibinfo {author}
  {\bibfnamefont {B.~H.}\ \bibnamefont {Kim}}, \bibinfo {author} {\bibfnamefont
  {S.-I.}\ \bibnamefont {Park}}, \bibinfo {author} {\bibfnamefont
  {B.}~\bibnamefont {Panilaitis}}, \bibinfo {author} {\bibfnamefont
  {R.}~\bibnamefont {Ghaffari}}, \bibinfo {author} {\bibfnamefont
  {J.}~\bibnamefont {Yao}}, \bibinfo {author} {\bibfnamefont {M.}~\bibnamefont
  {Li}}, \bibinfo {author} {\bibfnamefont {Z.}~\bibnamefont {Liu}}, \bibinfo
  {author} {\bibfnamefont {V.}~\bibnamefont {Malyarchuk}}, \bibinfo {author}
  {\bibfnamefont {D.~G.}\ \bibnamefont {Kim}}, \bibinfo {author} {\bibfnamefont
  {A.-P.}\ \bibnamefont {Le}}, \bibinfo {author} {\bibfnamefont {R.~G.}\
  \bibnamefont {Nuzzo}}, \bibinfo {author} {\bibfnamefont {D.~L.}\ \bibnamefont
  {Kaplan}}, \bibinfo {author} {\bibfnamefont {F.~G.}\ \bibnamefont
  {Omenetto}}, \bibinfo {author} {\bibfnamefont {Y.}~\bibnamefont {Huang}},
  \bibinfo {author} {\bibfnamefont {Z.}~\bibnamefont {Kang}}, \ and\ \bibinfo
  {author} {\bibfnamefont {J.~A.}\ \bibnamefont {Rogers}},\ }
  {\bibfield  {journal} {\bibinfo  {journal} {Nat. Materials}\ }\textbf
  {\bibinfo {volume} {9}},\ \bibinfo {pages} {929} (\bibinfo {year}
  {2010})}\BibitemShut {NoStop}%
\bibitem [{\citenamefont {Bowden}\ \emph {et~al.}(1998)\citenamefont {Bowden},
  \citenamefont {Brittain}, \citenamefont {Evans}, \citenamefont {Hutchinson},\
  and\ \citenamefont {Whitesides}}]{Bowden1998}%
  \BibitemOpen
  \bibfield  {author} {\bibinfo {author} {\bibfnamefont {N.}~\bibnamefont
  {Bowden}}, \bibinfo {author} {\bibfnamefont {S.}~\bibnamefont {Brittain}},
  \bibinfo {author} {\bibfnamefont {A.~G.}\ \bibnamefont {Evans}}, \bibinfo
  {author} {\bibfnamefont {J.~W.}\ \bibnamefont {Hutchinson}}, \ and\ \bibinfo
  {author} {\bibfnamefont {G.~M.}\ \bibnamefont {Whitesides}},\ } {\bibfield  {journal} {\bibinfo  {journal}
  {Nature}\ }\textbf {\bibinfo {volume} {393}},\ \bibinfo {pages} {146}
  (\bibinfo {year} {1998})}\BibitemShut {NoStop}%
\bibitem [{\citenamefont {Schweikart}\ and\ \citenamefont
  {Fery}(2009)}]{Schweikart2009}%
  \BibitemOpen
  \bibfield  {author} {\bibinfo {author} {\bibfnamefont {A.}~\bibnamefont
  {Schweikart}}\ and\ \bibinfo {author} {\bibfnamefont {A.}~\bibnamefont
  {Fery}},\ }\href {\doibase 10.1007/s00604-009-0153-3} {\bibfield  {journal}
  {\bibinfo  {journal} {Microchim. Acta}\ }\textbf {\bibinfo {volume} {165}},\
  \bibinfo {pages} {249} (\bibinfo {year} {2009})}\BibitemShut {NoStop}%
\bibitem [{\citenamefont {Kim}\ \emph {et~al.}(2012)\citenamefont {Kim},
  \citenamefont {Kim}, \citenamefont {Pegard}, \citenamefont {Oh},
  \citenamefont {Kagan}, \citenamefont {Fleischer}, \citenamefont {Stone},\
  and\ \citenamefont {Loo}}]{Kim2012b}%
  \BibitemOpen
  \bibfield  {author} {\bibinfo {author} {\bibfnamefont {J.~B.}\ \bibnamefont
  {Kim}}, \bibinfo {author} {\bibfnamefont {P.}~\bibnamefont {Kim}}, \bibinfo
  {author} {\bibfnamefont {N.~C.}\ \bibnamefont {Pegard}}, \bibinfo {author}
  {\bibfnamefont {S.~J.}\ \bibnamefont {Oh}}, \bibinfo {author} {\bibfnamefont
  {C.~R.}\ \bibnamefont {Kagan}}, \bibinfo {author} {\bibfnamefont {J.~W.}\
  \bibnamefont {Fleischer}}, \bibinfo {author} {\bibfnamefont {H.~A.}\
  \bibnamefont {Stone}}, \ and\ \bibinfo {author} {\bibfnamefont {Y.-L.}\
  \bibnamefont {Loo}},\ }
  {\bibfield  {journal} {\bibinfo  {journal} {Nat. Photonics}\ }\textbf
  {\bibinfo {volume} {6}},\ \bibinfo {pages} {327} (\bibinfo {year}
  {2012})}\BibitemShut {NoStop}%
\bibitem [{\citenamefont {Lin}\ \emph {et~al.}(2015)\citenamefont {Lin},
  \citenamefont {Wang}, \citenamefont {Gan}, \citenamefont {Hou}, \citenamefont
  {Yin},\ and\ \citenamefont {Jiang}}]{Lin2015}%
  \BibitemOpen
  \bibfield  {author} {\bibinfo {author} {\bibfnamefont {H.}~\bibnamefont
  {Lin}}, \bibinfo {author} {\bibfnamefont {Y.}~\bibnamefont {Wang}}, \bibinfo
  {author} {\bibfnamefont {Y.}~\bibnamefont {Gan}}, \bibinfo {author}
  {\bibfnamefont {H.}~\bibnamefont {Hou}}, \bibinfo {author} {\bibfnamefont
  {J.}~\bibnamefont {Yin}}, \ and\ \bibinfo {author} {\bibfnamefont
  {X.}~\bibnamefont {Jiang}},\ }\href {\doibase 10.1021/acs.langmuir.5b03484}
  {\bibfield  {journal} {\bibinfo  {journal} {Langmuir}\ }\textbf {\bibinfo
  {volume} {31}},\ \bibinfo {pages} {11800} (\bibinfo {year}
  {2015})}\BibitemShut {NoStop}%
\bibitem [{\citenamefont {Cerda}\ and\ \citenamefont
  {Mahadevan}(2003)}]{Cerda2003}%
  \BibitemOpen
  \bibfield  {author} {\bibinfo {author} {\bibfnamefont {E.}~\bibnamefont
  {Cerda}}\ and\ \bibinfo {author} {\bibfnamefont {L.}~\bibnamefont
  {Mahadevan}},\ }\href {\doibase 10.1103/PhysRevLett.90.074302} {\bibfield
  {journal} {\bibinfo  {journal} {Phys. Rev. Lett.}\ }\textbf {\bibinfo
  {volume} {90}},\ \bibinfo {pages} {074302} (\bibinfo {year}
  {2003})}\BibitemShut {NoStop}%
\bibitem [{\citenamefont {Li}\ \emph {et~al.}(2012)\citenamefont {Li},
  \citenamefont {Cao}, \citenamefont {Feng},\ and\ \citenamefont
  {Gao}}]{Bo2012}%
  \BibitemOpen
  \bibfield  {author} {\bibinfo {author} {\bibfnamefont {B.}~\bibnamefont
  {Li}}, \bibinfo {author} {\bibfnamefont {Y.-P.}\ \bibnamefont {Cao}},
  \bibinfo {author} {\bibfnamefont {X.-Q.}\ \bibnamefont {Feng}}, \ and\
  \bibinfo {author} {\bibfnamefont {H.}~\bibnamefont {Gao}},\ }\href {\doibase
  10.1039/C2SM00011C} {\bibfield  {journal} {\bibinfo  {journal} {Soft Matter}\
  }\textbf {\bibinfo {volume} {8}},\ \bibinfo {pages} {5728} (\bibinfo {year}
  {2012})}\BibitemShut {NoStop}%
\bibitem [{\citenamefont {Reis}\ \emph {et~al.}(2009)\citenamefont {Reis},
  \citenamefont {Corson}, \citenamefont {Boudaoud},\ and\ \citenamefont
  {Roman}}]{Reis2009}%
  \BibitemOpen
  \bibfield  {author} {\bibinfo {author} {\bibfnamefont {P.}~\bibnamefont
  {Reis}}, \bibinfo {author} {\bibfnamefont {F.}~\bibnamefont {Corson}},
  \bibinfo {author} {\bibfnamefont {A.}~\bibnamefont {Boudaoud}}, \ and\
  \bibinfo {author} {\bibfnamefont {B.}~\bibnamefont {Roman}},\ }\href
  {\doibase 10.1103/PhysRevLett.103.045501} {\bibfield  {journal} {\bibinfo
  {journal} {Phys. Rev. Lett.}\ }\textbf {\bibinfo {volume} {103}},\ \bibinfo
  {pages} {045501} (\bibinfo {year} {2009})}\BibitemShut {NoStop}%
\bibitem [{\citenamefont {Brau}\ \emph {et~al.}(2011)\citenamefont {Brau},
  \citenamefont {Vandeparre}, \citenamefont {Sabbah}, \citenamefont {Poulard},
  \citenamefont {Boudaoud},\ and\ \citenamefont {Damman}}]{Brau2011}%
  \BibitemOpen
  \bibfield  {author} {\bibinfo {author} {\bibfnamefont {F.}~\bibnamefont
  {Brau}}, \bibinfo {author} {\bibfnamefont {H.}~\bibnamefont {Vandeparre}},
  \bibinfo {author} {\bibfnamefont {A.}~\bibnamefont {Sabbah}}, \bibinfo
  {author} {\bibfnamefont {C.}~\bibnamefont {Poulard}}, \bibinfo {author}
  {\bibfnamefont {A.}~\bibnamefont {Boudaoud}}, \ and\ \bibinfo {author}
  {\bibfnamefont {P.}~\bibnamefont {Damman}},\ }
  {\bibfield  {journal} {\bibinfo  {journal} {Nat. Phys.}\ }\textbf {\bibinfo
  {volume} {7}},\ \bibinfo {pages} {56} (\bibinfo {year} {2011})}\BibitemShut
  {NoStop}%
\bibitem [{\citenamefont {Kim}\ \emph {et~al.}(2011)\citenamefont {Kim},
  \citenamefont {Abkarian},\ and\ \citenamefont {Stone}}]{Kim2011}%
  \BibitemOpen
  \bibfield  {author} {\bibinfo {author} {\bibfnamefont {P.}~\bibnamefont
  {Kim}}, \bibinfo {author} {\bibfnamefont {M.}~\bibnamefont {Abkarian}}, \
  and\ \bibinfo {author} {\bibfnamefont {H.~A.}\ \bibnamefont {Stone}},\ }
  {\bibfield  {journal} {\bibinfo  {journal} {Nat. Materials}\ }\textbf
  {\bibinfo {volume} {10}},\ \bibinfo {pages} {952} (\bibinfo {year}
  {2011})}\BibitemShut {NoStop}%
\bibitem [{\citenamefont {Levien}(2008)}]{Levien2008}%
  \BibitemOpen
  \bibfield  {author} {\bibinfo {author} {\bibfnamefont {R.}~\bibnamefont
  {Levien}},\ }\href@noop {} {\emph {\bibinfo {title} {The {E}lastica: {A}
  Mathematical History}}},\ \bibinfo {type} {Tech. Rep.}\ (\bibinfo
  {institution} {{EECS Department}, {U}niversity of {C}alifornia, {B}erkeley},\
  \bibinfo {year} {2008})\BibitemShut {NoStop}%
\bibitem [{\citenamefont {Landau}\ and\ \citenamefont
  {Lifschitz}(1970)}]{Landau}%
  \BibitemOpen
  \bibfield  {author} {\bibinfo {author} {\bibfnamefont {L.~D.}\ \bibnamefont
  {Landau}}\ and\ \bibinfo {author} {\bibfnamefont {E.~M.}\ \bibnamefont
  {Lifschitz}},\ }\href@noop {} {\emph {\bibinfo {title} {The Theory of
  Elasticity}}}\ (\bibinfo  {publisher} {Pergamon},\ \bibinfo {year}
  {1970})\BibitemShut {NoStop}%
\bibitem [{\citenamefont {Pocivavsek}\ \emph {et~al.}(2008)\citenamefont
  {Pocivavsek}, \citenamefont {Dellsy}, \citenamefont {Kern}, \citenamefont
  {Johnson}, \citenamefont {Lin}, \citenamefont {Lee},\ and\ \citenamefont
  {Cerda}}]{Pocivavsek2008}%
  \BibitemOpen
  \bibfield  {author} {\bibinfo {author} {\bibfnamefont {L.}~\bibnamefont
  {Pocivavsek}}, \bibinfo {author} {\bibfnamefont {R.}~\bibnamefont {Dellsy}},
  \bibinfo {author} {\bibfnamefont {A.}~\bibnamefont {Kern}}, \bibinfo {author}
  {\bibfnamefont {S.}~\bibnamefont {Johnson}}, \bibinfo {author} {\bibfnamefont
  {B.}~\bibnamefont {Lin}}, \bibinfo {author} {\bibfnamefont {K.~Y.~C.}\
  \bibnamefont {Lee}}, \ and\ \bibinfo {author} {\bibfnamefont
  {E.}~\bibnamefont {Cerda}},\ }\href {\doibase 10.1126/science.1154069}
  {\bibfield  {journal} {\bibinfo  {journal} {Science}\ }\textbf {\bibinfo
  {volume} {320}},\ \bibinfo {pages} {912} (\bibinfo {year}
  {2008})}\BibitemShut {NoStop}%
\bibitem [{\citenamefont {Holmes}\ and\ \citenamefont
  {Crosby}(2010)}]{Holmes2010}%
  \BibitemOpen
  \bibfield  {author} {\bibinfo {author} {\bibfnamefont {D.~P.}\ \bibnamefont
  {Holmes}}\ and\ \bibinfo {author} {\bibfnamefont {A.~J.}\ \bibnamefont
  {Crosby}},\ }\href {\doibase 10.1103/PhysRevLett.105.038303} {\bibfield
  {journal} {\bibinfo  {journal} {Phys. Rev. Lett.}\ }\textbf {\bibinfo
  {volume} {105}},\ \bibinfo {pages} {038303} (\bibinfo {year}
  {2010})}\BibitemShut {NoStop}%
\bibitem [{\citenamefont {King}\ \emph {et~al.}(2012)\citenamefont {King},
  \citenamefont {Schroll}, \citenamefont {Davidovitch},\ and\ \citenamefont
  {Menon}}]{King2012}%
  \BibitemOpen
  \bibfield  {author} {\bibinfo {author} {\bibfnamefont {H.}~\bibnamefont
  {King}}, \bibinfo {author} {\bibfnamefont {R.~D.}\ \bibnamefont {Schroll}},
  \bibinfo {author} {\bibfnamefont {B.}~\bibnamefont {Davidovitch}}, \ and\
  \bibinfo {author} {\bibfnamefont {N.}~\bibnamefont {Menon}},\ }\href
  {\doibase 10.1073/pnas.1201201109} {\bibfield  {journal} {\bibinfo  {journal}
  {Proc. Natl. Acad. Sci. U. S. A.}\ }\textbf {\bibinfo {volume} {109}},\
  \bibinfo {pages} {9716} (\bibinfo {year} {2012})}\BibitemShut {NoStop}%
\bibitem [{\citenamefont {Pi{\~{n}}eirua}\ \emph {et~al.}(2013)\citenamefont
  {Pi{\~{n}}eirua}, \citenamefont {Tanaka}, \citenamefont {Roman},\ and\
  \citenamefont {Bico}}]{Pineirua2013}%
  \BibitemOpen
  \bibfield  {author} {\bibinfo {author} {\bibfnamefont {M.}~\bibnamefont
  {Pi{\~{n}}eirua}}, \bibinfo {author} {\bibfnamefont {N.}~\bibnamefont
  {Tanaka}}, \bibinfo {author} {\bibfnamefont {B.}~\bibnamefont {Roman}}, \
  and\ \bibinfo {author} {\bibfnamefont {J.}~\bibnamefont {Bico}},\ }\href
  {\doibase 10.1039/c3sm51825f} {\bibfield  {journal} {\bibinfo  {journal}
  {Soft Matter}\ }\textbf {\bibinfo {volume} {9}},\ \bibinfo {pages} {10985}
  (\bibinfo {year} {2013})}\BibitemShut {NoStop}%
\bibitem [{\citenamefont {Leahy}\ \emph {et~al.}(2010)\citenamefont {Leahy},
  \citenamefont {Pocivavsek}, \citenamefont {Meron}, \citenamefont {Lam},
  \citenamefont {Salas}, \citenamefont {Viccaro}, \citenamefont {Lee},\ and\
  \citenamefont {Lin}}]{Leahy2010}%
  \BibitemOpen
  \bibfield  {author} {\bibinfo {author} {\bibfnamefont {B.~D.}\ \bibnamefont
  {Leahy}}, \bibinfo {author} {\bibfnamefont {L.}~\bibnamefont {Pocivavsek}},
  \bibinfo {author} {\bibfnamefont {M.}~\bibnamefont {Meron}}, \bibinfo
  {author} {\bibfnamefont {K.~L.}\ \bibnamefont {Lam}}, \bibinfo {author}
  {\bibfnamefont {D.}~\bibnamefont {Salas}}, \bibinfo {author} {\bibfnamefont
  {P.~J.}\ \bibnamefont {Viccaro}}, \bibinfo {author} {\bibfnamefont
  {K.~Y.~C.}\ \bibnamefont {Lee}}, \ and\ \bibinfo {author} {\bibfnamefont
  {B.}~\bibnamefont {Lin}},\ }\href {\doibase 10.1103/PhysRevLett.105.058301}
  {\bibfield  {journal} {\bibinfo  {journal} {Phys. Rev. Lett.}\ }\textbf
  {\bibinfo {volume} {105}},\ \bibinfo {pages} {058301} (\bibinfo {year}
  {2010})}\BibitemShut {NoStop}%
\bibitem [{\citenamefont {Lu}\ \emph {et~al.}(2002)\citenamefont {Lu},
  \citenamefont {Knobler}, \citenamefont {Bruinsma}, \citenamefont {Twardos},\
  and\ \citenamefont {Dennin}}]{Lu2002}%
  \BibitemOpen
  \bibfield  {author} {\bibinfo {author} {\bibfnamefont {W.}~\bibnamefont
  {Lu}}, \bibinfo {author} {\bibfnamefont {C.}~\bibnamefont {Knobler}},
  \bibinfo {author} {\bibfnamefont {R.}~\bibnamefont {Bruinsma}}, \bibinfo
  {author} {\bibfnamefont {M.}~\bibnamefont {Twardos}}, \ and\ \bibinfo
  {author} {\bibfnamefont {M.}~\bibnamefont {Dennin}},\ }\href {\doibase
  10.1103/PhysRevLett.89.146107} {\bibfield  {journal} {\bibinfo  {journal}
  {Phys. Rev. Lett.}\ }\textbf {\bibinfo {volume} {89}},\ \bibinfo {pages}
  {146107} (\bibinfo {year} {2002})}\BibitemShut {NoStop}%
\bibitem [{\citenamefont {Boatwright}\ \emph {et~al.}(2010)\citenamefont
  {Boatwright}, \citenamefont {Levine},\ and\ \citenamefont
  {Dennin}}]{Boatwright2010}%
  \BibitemOpen
  \bibfield  {author} {\bibinfo {author} {\bibfnamefont {T.}~\bibnamefont
  {Boatwright}}, \bibinfo {author} {\bibfnamefont {A.~J.}\ \bibnamefont
  {Levine}}, \ and\ \bibinfo {author} {\bibfnamefont {M.}~\bibnamefont
  {Dennin}},\ }\href {\doibase 10.1021/la1012439} {\bibfield  {journal}
  {\bibinfo  {journal} {Langmuir}\ }\textbf {\bibinfo {volume} {26}},\ \bibinfo
  {pages} {12755} (\bibinfo {year} {2010})}\BibitemShut {NoStop}%
\bibitem [{\citenamefont {Vella}\ \emph {et~al.}(2004)\citenamefont {Vella},
  \citenamefont {Aussillous},\ and\ \citenamefont {Mahadevan}}]{Vella2004}%
  \BibitemOpen
  \bibfield  {author} {\bibinfo {author} {\bibfnamefont {D.}~\bibnamefont
  {Vella}}, \bibinfo {author} {\bibfnamefont {P.}~\bibnamefont {Aussillous}}, \
  and\ \bibinfo {author} {\bibfnamefont {L.}~\bibnamefont {Mahadevan}},\ }{\bibfield  {journal}
  {\bibinfo  {journal} {EPL (Europhysics Letters)}\ }\textbf {\bibinfo {volume}
  {68}},\ \bibinfo {pages} {212} (\bibinfo {year} {2004})}\BibitemShut
  {NoStop}%
\bibitem [{\citenamefont {Wagner}\ and\ \citenamefont
  {Vella}(2011)}]{Wagner2011}%
  \BibitemOpen
  \bibfield  {author} {\bibinfo {author} {\bibfnamefont {T.~J.~W.}\
  \bibnamefont {Wagner}}\ and\ \bibinfo {author} {\bibfnamefont
  {D.}~\bibnamefont {Vella}},\ }\href@noop {} {\bibfield  {journal} {\bibinfo
  {journal} {Phys. Rev. Lett.}\ }\textbf {\bibinfo {volume} {107}},\ \bibinfo
  {pages} {044301} (\bibinfo {year} {2011})}\BibitemShut {NoStop}%
\bibitem [{\citenamefont {Diamant}\ and\ \citenamefont
  {Witten}(2011)}]{Diamant2011}%
  \BibitemOpen
  \bibfield  {author} {\bibinfo {author} {\bibfnamefont {H.}~\bibnamefont
  {Diamant}}\ and\ \bibinfo {author} {\bibfnamefont {T.~A.}\ \bibnamefont
  {Witten}},\ }\href {\doibase 10.1103/PhysRevLett.107.164302} {\bibfield
  {journal} {\bibinfo  {journal} {Phys. Rev. Lett.}\ }\textbf {\bibinfo
  {volume} {107}},\ \bibinfo {pages} {164302} (\bibinfo {year}
  {2011})}\BibitemShut {NoStop}%
\bibitem [{\citenamefont {Diamant}\ and\ \citenamefont
  {Witten}(2013)}]{Diamant2013}%
  \BibitemOpen
  \bibfield  {author} {\bibinfo {author} {\bibfnamefont {H.}~\bibnamefont
  {Diamant}}\ and\ \bibinfo {author} {\bibfnamefont {T.~A.}\ \bibnamefont
  {Witten}},\ }\href {\doibase 10.1103/PhysRevE.88.012401} {\bibfield
  {journal} {\bibinfo  {journal} {Phys. Rev. E}\ }\textbf {\bibinfo {volume}
  {88}},\ \bibinfo {pages} {012401} (\bibinfo {year} {2013})}\BibitemShut
  {NoStop}%
\bibitem [{\citenamefont {Brau}\ \emph {et~al.}(2013)\citenamefont {Brau},
  \citenamefont {Damman}, \citenamefont {Diamant},\ and\ \citenamefont
  {Witten}}]{Brau2013}%
  \BibitemOpen
  \bibfield  {author} {\bibinfo {author} {\bibfnamefont {F.}~\bibnamefont
  {Brau}}, \bibinfo {author} {\bibfnamefont {P.}~\bibnamefont {Damman}},
  \bibinfo {author} {\bibfnamefont {H.}~\bibnamefont {Diamant}}, \ and\
  \bibinfo {author} {\bibfnamefont {T.~A.}\ \bibnamefont {Witten}},\
  }\href@noop {} {\bibfield  {journal} {\bibinfo  {journal} {Soft Matter}\
  }\textbf {\bibinfo {volume} {9}},\ \bibinfo {pages} {8177} (\bibinfo {year}
  {2013})}\BibitemShut {NoStop}%
\bibitem [{\citenamefont {Rivetti}(2013)}]{Rivetti2013}%
  \BibitemOpen
  \bibfield  {author} {\bibinfo {author} {\bibfnamefont {M.}~\bibnamefont
  {Rivetti}},\ }\href {\doibase 10.1016/j.crme.2013.01.005} {\bibfield
  {journal} {\bibinfo  {journal} {C. R. Mecanique}\ }\textbf {\bibinfo {volume}
  {341}},\ \bibinfo {pages} {333} (\bibinfo {year} {2013})}\BibitemShut
  {NoStop}%
\bibitem [{\citenamefont {Rivetti}\ and\ \citenamefont
  {Neukirch}(2014)}]{Rivetti2014}%
  \BibitemOpen
  \bibfield  {author} {\bibinfo {author} {\bibfnamefont {M.}~\bibnamefont
  {Rivetti}}\ and\ \bibinfo {author} {\bibfnamefont {S.}~\bibnamefont
  {Neukirch}},\ }\href {\doibase 10.1016/j.jmps.2014.05.004} {\bibfield
  {journal} {\bibinfo  {journal} {J. Mech. Phys. Solids}\ }\textbf {\bibinfo
  {volume} {69}},\ \bibinfo {pages} {143} (\bibinfo {year} {2014})}\BibitemShut
  {NoStop}%
\bibitem [{\citenamefont {Oshri}\ \emph {et~al.}(2015)\citenamefont {Oshri},
  \citenamefont {Brau},\ and\ \citenamefont {Diamant}}]{Oshri2015}%
  \BibitemOpen
  \bibfield  {author} {\bibinfo {author} {\bibfnamefont {O.}~\bibnamefont
  {Oshri}}, \bibinfo {author} {\bibfnamefont {F.}~\bibnamefont {Brau}}, \ and\
  \bibinfo {author} {\bibfnamefont {H.}~\bibnamefont {Diamant}},\ }\href
  {\doibase 10.1103/PhysRevE.91.052408} {\bibfield  {journal} {\bibinfo
  {journal} {Phys. Rev. E}\ }\textbf {\bibinfo {volume} {91}},\ \bibinfo
  {pages} {052408} (\bibinfo {year} {2015})}\BibitemShut {NoStop}%
\bibitem [{\citenamefont {D{\'{e}}mery}\ \emph {et~al.}(2014)\citenamefont
  {D{\'{e}}mery}, \citenamefont {Davidovitch},\ and\ \citenamefont
  {Santangelo}}]{Demery2014}%
  \BibitemOpen
  \bibfield  {author} {\bibinfo {author} {\bibfnamefont {V.}~\bibnamefont
  {D{\'{e}}mery}}, \bibinfo {author} {\bibfnamefont {B.}~\bibnamefont
  {Davidovitch}}, \ and\ \bibinfo {author} {\bibfnamefont {C.~D.}\ \bibnamefont
  {Santangelo}},\ }\href {\doibase 10.1103/PhysRevE.90.042401} {\bibfield
  {journal} {\bibinfo  {journal} {Phys. Rev. E}\ }\textbf {\bibinfo {volume}
  {90}},\ \bibinfo {pages} {042401} (\bibinfo {year} {2014})}
  \BibitemShut {NoStop}%
\bibitem [{\citenamefont {{Wang}}\ and\ \citenamefont
  {{Drachman}}(1981)}]{Wang1981}%
  \BibitemOpen
  \bibfield  {author} {\bibinfo {author} {\bibfnamefont {C.-Y.}\ \bibnamefont
  {{Wang}}}\ and\ \bibinfo {author} {\bibfnamefont {B.}~\bibnamefont
  {{Drachman}}},\ }\href {\doibase 10.1115/1.3157696} {\bibfield  {journal}
  {\bibinfo  {journal} {J. Appl. Mech.}\ }\textbf {\bibinfo {volume} {48}},\
  \bibinfo {pages} {668} (\bibinfo {year} {1981})}\BibitemShut {NoStop}%
\bibitem [{\citenamefont {Wang}(1986)}]{Wang1986}%
  \BibitemOpen
  \bibfield  {author} {\bibinfo {author} {\bibfnamefont {C.}~\bibnamefont
  {Wang}},\ }
  {\bibfield  {journal} {\bibinfo  {journal} {Int. J. Mech. Sci.}\ }\textbf
  {\bibinfo {volume} {28}},\ \bibinfo {pages} {549 } (\bibinfo {year}
  {1986})}\BibitemShut {NoStop}%
\bibitem [{\citenamefont {Bosi}\ \emph {et~al.}(2015)\citenamefont {Bosi},
  \citenamefont {Misseroni}, \citenamefont {Dal~Corso},\ and\ \citenamefont
  {Bigoni}}]{Bosi2015}%
  \BibitemOpen
  \bibfield  {author} {\bibinfo {author} {\bibfnamefont {F.}~\bibnamefont
  {Bosi}}, \bibinfo {author} {\bibfnamefont {D.}~\bibnamefont {Misseroni}},
  \bibinfo {author} {\bibfnamefont {F.}~\bibnamefont {Dal~Corso}}, \ and\
  \bibinfo {author} {\bibfnamefont {D.}~\bibnamefont {Bigoni}},\ }\href
  {\doibase 10.1098/rspa.2015.01952} {\bibfield  {journal} {\bibinfo  {journal}
  {Proc. R. Soc. A}\ }\textbf {\bibinfo {volume} {471}},\ \bibinfo {pages}
  {20150195} (\bibinfo {year} {2015})}\BibitemShut {NoStop}%
\bibitem [{\citenamefont {Subramaniam}\ \emph {et~al.}(2005)\citenamefont
  {Subramaniam}, \citenamefont {Abkarian}, \citenamefont {Mahadevan},\ and\
  \citenamefont {Stone}}]{Subramaniam2005}%
  \BibitemOpen
  \bibfield  {author} {\bibinfo {author} {\bibfnamefont {A.~B.}\ \bibnamefont
  {Subramaniam}}, \bibinfo {author} {\bibfnamefont {M.}~\bibnamefont
  {Abkarian}}, \bibinfo {author} {\bibfnamefont {L.}~\bibnamefont {Mahadevan}},
  \ and\ \bibinfo {author} {\bibfnamefont {H.~A.}\ \bibnamefont {Stone}},\
  }\href@noop {} {\bibfield  {journal} {\bibinfo  {journal} {Nature}\ }\textbf
  {\bibinfo {volume} {438}},\ \bibinfo {pages} {930} (\bibinfo {year}
  {2005})}\BibitemShut {NoStop}%
\bibitem [{\citenamefont {Petit}\ \emph {et~al.}(2016)\citenamefont {Petit},
  \citenamefont {Biance}, \citenamefont {Lorenceau},\ and\ \citenamefont
  {Planchette}}]{Petit2016}%
  \BibitemOpen
  \bibfield  {author} {\bibinfo {author} {\bibfnamefont {P.}~\bibnamefont
  {Petit}}, \bibinfo {author} {\bibfnamefont {A.-L.}\ \bibnamefont {Biance}},
  \bibinfo {author} {\bibfnamefont {E.}~\bibnamefont {Lorenceau}}, \ and\
  \bibinfo {author} {\bibfnamefont {C.}~\bibnamefont {Planchette}},\
  }\href@noop {} {\bibfield  {journal} {\bibinfo  {journal} {Physical Review
  E}\ }\textbf {\bibinfo {volume} {93}},\ \bibinfo {pages} {042802} (\bibinfo
  {year} {2016})}\BibitemShut {NoStop}%
\bibitem [{\citenamefont {Cicuta}\ and\ \citenamefont
  {Vella}(2009)}]{Cicuta2009}%
  \BibitemOpen
  \bibfield  {author} {\bibinfo {author} {\bibfnamefont {P.}~\bibnamefont
  {Cicuta}}\ and\ \bibinfo {author} {\bibfnamefont {D.}~\bibnamefont {Vella}},\
  }\href@noop {} {\bibfield  {journal} {\bibinfo  {journal} {Physical Review
  Letters}\ }\textbf {\bibinfo {volume} {102}},\ \bibinfo {pages} {138302}
  (\bibinfo {year} {2009})}\BibitemShut {NoStop}%
\end{thebibliography}

\begin{thebibliography}{7}%
\makeatletter
\providecommand \@ifxundefined [1]{%
 \@ifx{#1\undefined}
}%
\providecommand \@ifnum [1]{%
 \ifnum #1\expandafter \@firstoftwo
 \else \expandafter \@secondoftwo
 \fi
}%
\providecommand \@ifx [1]{%
 \ifx #1\expandafter \@firstoftwo
 \else \expandafter \@secondoftwo
 \fi
}%
\providecommand \natexlab [1]{#1}%
\providecommand \enquote  [1]{``#1''}%
\providecommand \bibnamefont  [1]{#1}%
\providecommand \bibfnamefont [1]{#1}%
\providecommand \citenamefont [1]{#1}%
\providecommand \href@noop [0]{\@secondoftwo}%
\providecommand \href [0]{\begingroup \@sanitize@url \@href}%
\providecommand \@href[1]{\@@startlink{#1}\@@href}%
\providecommand \@@href[1]{\endgroup#1\@@endlink}%
\providecommand \@sanitize@url [0]{\catcode `\\12\catcode `\$12\catcode
  `\&12\catcode `\#12\catcode `\^12\catcode `\_12\catcode `\%12\relax}%
\providecommand \@@startlink[1]{}%
\providecommand \@@endlink[0]{}%
\providecommand \url  [0]{\begingroup\@sanitize@url \@url }%
\providecommand \@url [1]{\endgroup\@href {#1}{\urlprefix }}%
\providecommand \urlprefix  [0]{URL }%
\providecommand \Eprint [0]{\href }%
\providecommand \doibase [0]{http://dx.doi.org/}%
\providecommand \selectlanguage [0]{\@gobble}%
\providecommand \bibinfo  [0]{\@secondoftwo}%
\providecommand \bibfield  [0]{\@secondoftwo}%
\providecommand \translation [1]{[#1]}%
\providecommand \BibitemOpen [0]{}%
\providecommand \bibitemStop [0]{}%
\providecommand \bibitemNoStop [0]{.\EOS\space}%
\providecommand \EOS [0]{\spacefactor3000\relax}%
\providecommand \BibitemShut  [1]{\csname bibitem#1\endcsname}%
\let\auto@bib@innerbib\@empty
\bibitem [{\citenamefont {Diamant}\ and\ \citenamefont
  {Witten}(2011)}]{Diamant2011SI}%
  \BibitemOpen
  \bibfield  {author} {\bibinfo {author} {\bibfnamefont {H.}~\bibnamefont
  {Diamant}}\ and\ \bibinfo {author} {\bibfnamefont {T.~A.}\ \bibnamefont
  {Witten}},\ }\href {\doibase 10.1103/PhysRevLett.107.164302} {\bibfield
  {journal} {\bibinfo  {journal} {Phys. Rev. Lett.}\ }\textbf {\bibinfo
  {volume} {107}},\ \bibinfo {pages} {164302} (\bibinfo {year}
  {2011})}\BibitemShut {NoStop}%
\bibitem [{\citenamefont {Oshri}\ \emph {et~al.}(2015)\citenamefont {Oshri},
  \citenamefont {Brau},\ and\ \citenamefont {Diamant}}]{Oshri2015SI}%
  \BibitemOpen
  \bibfield  {author} {\bibinfo {author} {\bibfnamefont {O.}~\bibnamefont
  {Oshri}}, \bibinfo {author} {\bibfnamefont {F.}~\bibnamefont {Brau}}, \ and\
  \bibinfo {author} {\bibfnamefont {H.}~\bibnamefont {Diamant}},\ }\href
  {\doibase 10.1103/PhysRevE.91.052408} {\bibfield  {journal} {\bibinfo
  {journal} {Phys. Rev. E}\ }\textbf {\bibinfo {volume} {91}},\ \bibinfo
  {pages} {052408} (\bibinfo {year} {2015})}\BibitemShut {NoStop}%
\bibitem [{\citenamefont {Pocivavsek}\ \emph {et~al.}(2008)\citenamefont
  {Pocivavsek}, \citenamefont {Dellsy}, \citenamefont {Kern}, \citenamefont
  {Johnson}, \citenamefont {Lin}, \citenamefont {Lee},\ and\ \citenamefont
  {Cerda}}]{Pocivavsek2008SI}%
  \BibitemOpen
  \bibfield  {author} {\bibinfo {author} {\bibfnamefont {L.}~\bibnamefont
  {Pocivavsek}}, \bibinfo {author} {\bibfnamefont {R.}~\bibnamefont {Dellsy}},
  \bibinfo {author} {\bibfnamefont {A.}~\bibnamefont {Kern}}, \bibinfo {author}
  {\bibfnamefont {S.}~\bibnamefont {Johnson}}, \bibinfo {author} {\bibfnamefont
  {B.}~\bibnamefont {Lin}}, \bibinfo {author} {\bibfnamefont {K.~Y.~C.}\
  \bibnamefont {Lee}}, \ and\ \bibinfo {author} {\bibfnamefont
  {E.}~\bibnamefont {Cerda}},\ }\href {\doibase 10.1126/science.1154069}
  {\bibfield  {journal} {\bibinfo  {journal} {Science}\ }\textbf {\bibinfo
  {volume} {320}},\ \bibinfo {pages} {912} (\bibinfo {year}
  {2008})}\BibitemShut {NoStop}%
\bibitem [{\citenamefont {Diamant}\ and\ \citenamefont
  {Witten}(2013)}]{Diamant2013SI}%
  \BibitemOpen
  \bibfield  {author} {\bibinfo {author} {\bibfnamefont {H.}~\bibnamefont
  {Diamant}}\ and\ \bibinfo {author} {\bibfnamefont {T.~A.}\ \bibnamefont
  {Witten}},\ }\href {\doibase 10.1103/PhysRevE.88.012401} {\bibfield
  {journal} {\bibinfo  {journal} {Phys. Rev. E}\ }\textbf {\bibinfo {volume}
  {88}},\ \bibinfo {pages} {012401} (\bibinfo {year} {2013})}\BibitemShut
  {NoStop}%
\bibitem [{\citenamefont {Rivetti}(2013)}]{Rivetti2013SI}%
  \BibitemOpen
  \bibfield  {author} {\bibinfo {author} {\bibfnamefont {M.}~\bibnamefont
  {Rivetti}},\ }\href {\doibase 10.1016/j.crme.2013.01.005} {\bibfield
  {journal} {\bibinfo  {journal} {C. R. Mecanique}\ }\textbf {\bibinfo {volume}
  {341}},\ \bibinfo {pages} {333} (\bibinfo {year} {2013})}\BibitemShut
  {NoStop}%
\bibitem [{\citenamefont {Rivetti}\ and\ \citenamefont
  {Neukirch}(2014)}]{Rivetti2014SI}%
  \BibitemOpen
  \bibfield  {author} {\bibinfo {author} {\bibfnamefont {M.}~\bibnamefont
  {Rivetti}}\ and\ \bibinfo {author} {\bibfnamefont {S.}~\bibnamefont
  {Neukirch}},\ }\href {\doibase 10.1016/j.jmps.2014.05.004} {\bibfield
  {journal} {\bibinfo  {journal} {J. Mech. Phys. Solids}\ }\textbf {\bibinfo
  {volume} {69}},\ \bibinfo {pages} {143} (\bibinfo {year} {2014})}\BibitemShut
  {NoStop}%
\bibitem [{\citenamefont {Huang}\ \emph {et~al.}(2010)\citenamefont {Huang},
  \citenamefont {Davidovitch}, \citenamefont {Santangelo}, \citenamefont
  {Russell},\ and\ \citenamefont {Menon}}]{Huang2010SI}%
  \BibitemOpen
  \bibfield  {author} {\bibinfo {author} {\bibfnamefont {J.}~\bibnamefont
  {Huang}}, \bibinfo {author} {\bibfnamefont {B.}~\bibnamefont {Davidovitch}},
  \bibinfo {author} {\bibfnamefont {C.~D.}\ \bibnamefont {Santangelo}},
  \bibinfo {author} {\bibfnamefont {T.~P.}\ \bibnamefont {Russell}}, \ and\
  \bibinfo {author} {\bibfnamefont {N.}~\bibnamefont {Menon}},\ }\href
  {\doibase 10.1103/PhysRevLett.105.038302} {\bibfield  {journal} {\bibinfo
  {journal} {Phys. Rev. Lett.}\ }\textbf {\bibinfo {volume} {105}},\ \bibinfo
  {pages} {2} (\bibinfo {year} {2010})},\ \Eprint
  {http://arxiv.org/abs/0901.2892} {arXiv:0901.2892} \BibitemShut {NoStop}%
\end{thebibliography}

%

\widetext
\clearpage
\begin{center}
\textbf{\large Supplementary information for ``\emph{The compression of a heavy floating elastic film}"}
\end{center}
\setcounter{equation}{0}
\setcounter{figure}{0}
\setcounter{table}{0}
\setcounter{section}{0}
\setcounter{page}{1}
\makeatletter
\renewcommand{\theequation}{S\arabic{equation}}
\renewcommand{\thesection}{S\arabic{section}}
\renewcommand{\thefigure}{S\arabic{figure}}
\renewcommand{\bibnumfmt}[1]{[S#1]}
\renewcommand{\citenumfont}[1]{S#1}

\section{Model of the floating heavy film prior to self-contact }
We consider an incompressible film of length $L_0$, width $W$, thickness $t$ and density $\rho_s$ lying between a fluid of density $\rho_{up}$ (for the upper fluid) and a lower liquid of density $\rho_{low}>\rho_{up}$. The film is compressed uniaxially in the $x$ direction (see Fig.~1). We introduce the intrinsic coordinates $(s,\theta)$ in which $s$ is the arclength and $\theta (s)$ is the local angle between the tangent and the horizontal axis $x$; we parametrize the film centreline in terms of arc-length, $[x(s),y(s)]$. 

Neglecting surface tension, the energy of this system is: $U=U_b + U_{low} + U_{up} + U_g$ with $U_b$ the bending energy, $U_{low}$ and $U_{up}$  the gravitational potential energies of the liquid/fluids that are displaced by the film and $U_g$ the  gravitational energy of the film itself. In our system of coordinates we have:
\begin{eqnarray}
U_b=\frac{BW}{2}\int_{-L_0/2}^{L_0/2} \left(\partial_s\theta\right)^2 \: \upd s \nonumber \\
U_{low} + U_{up}=\frac{\Delta \rho g W}{2}\int_{-L_0/2}^{L_0/2} y(s)^2 \cos \theta \: \upd s \nonumber \\
U_g=(\rho_s-\rho_{low})gWt \int_{-L_0/2}^{L_0/2} y(s)\: \upd s \nonumber
\end{eqnarray}
with $g$ the gravitational acceleration, $\Delta\rho=\rho_{low}-\rho_{up}$ the density difference between the fluids, $B$ the bending modulus (per unit width) of the beam, $B=Et^3/[12(1-\nu^2)]$, $E$ is the Young's modulus and $\nu$ the Poisson ratio. Assuming the film is inextensible, we have a global constraint on the end displacement:
\[\Delta = L_0-L=\int_{-L_0/2}^{L_0/2} (1- \cos \theta)\: \upd s \]
We also have a local constraint due to the use of intrinsic coordinates $\partial_s y (s)=\sin \theta(s)$.

To determine the equilibrium profile of the compressed film, we first minimize the total energy accounting for the two constraints mentioned above. This adds two Lagrange multipliers: $P$ for the end displacement (which corresponds physically  to the compressive force applied, $P=\partial_{\Delta}U$) and $Q(s)$ for the relation between $\partial_sy$ and $\theta$. To facilitate the calculation, we rescale lengths by $\ell_{eh}=(B/\Delta \rho g)^{1/4}$,  energy by $WB/\ell_{eh}$ and $P$ by $W(B \Delta \rho g)^{1/2}$ and only use dimensionless quantities in the following. We find that the energies are
\[U_b=\frac{1}{2}\int_{-L_0/2}^{L_0/2} \left(\partial_s\theta\right)^2\: \upd s, \quad U_{low} + U_{up}=\frac{1}{2}\int_{-L_0/2}^{L_0/2} y(s)^2 \cos \theta\: \upd s, \quad U_g=M \int_{-L_0/2}^{L_0/2} y(s)\: \upd s\]
In this process a  dimensionless number appears, 
\[M=\frac{(\rho_s-\rho_{low})t}{\Delta\rho \ell_{eh}},\]
which measures the weight of the film relative to the restoring force provided by Archimedes'  buoyancy over the horizontal length $\ell_{eh}$.
 The action to minimize is therefore $\mathcal{S}=\int_{-L_0/2}^{L_0/2} \upd s\: \mathcal{L}(\theta,\: \partial_s \theta,\: y,\: \partial_s y)$ with
\[\mathcal{L}=\tfrac{1}{2} \left(\partial_s\theta\right)^2 +\tfrac{1}{2}y^2 \cos \theta + My - P(1-\cos \theta) - Q(s)(\sin \theta - \partial_s y)\] 
We use Hamiltonian mechanics, following Diamant and Witten \cite{Diamant2011SI}, to perform the minimization. The conjugate momenta and the Hamiltonian are:
\begin{eqnarray*}
& p_\theta=\frac{\partial \mathcal{L}}{\partial (\partial_s \theta)}=\partial_s \theta, \quad p_y=\frac{\partial \mathcal{L}}{\partial (\partial_s y)}=Q \nonumber\\
& \mathcal{H}=p_\theta \partial_s \theta  + p_y \partial_s y -\mathcal{L} \nonumber
\end{eqnarray*}
Since $\mathcal{L}$ has no explicit dependence on $s$, $\mathcal{H}$ is a constant of motion, thus $\mathcal{H}(s)=\mathcal{H}(\pm L_0/2)$. 
Here we focus on localized deformations and therefore choose the boundary conditions
$$y(\pm L_0/2)=\theta (\pm L_0/2) = \partial_s \theta (\pm L_0/2) = 0,$$ which immediately gives that $\mathcal{H}(s)=0$, i.e.
\begin{equation}
\mathcal{H}=\tfrac{1}{2}\left(\partial_s\theta\right)^2+Q \sin \theta - \tfrac{1}{2}y^2 \cos \theta - My + P(1-\cos \theta)=0
\label{Heq}
\end{equation}
Hamilton's equations  $\partial_s p_\theta=-\partial \mathcal{H} / \partial \theta$ and $\partial_s p_y=-\partial \mathcal{H} / \partial y$ then give:
\begin{eqnarray}
\partial_s^2\theta+\left(\tfrac{1}{2}y^2+P\right)\sin\theta+Q\cos\theta=0 \label{firstmom}\\
\partial_sQ-y\cos\theta-M=0 \label{secmom}
\end{eqnarray}
If we differentiate \eqref{firstmom} with respect to $s$ and eliminate $Q\sin\theta$ with \eqref{Heq} and $\partial_sQ$ with \eqref{secmom} we get:
\begin{equation}
\partial_s^3\theta + \left[\tfrac{1}{2}\left(\partial_s\theta\right)^2+P\right]\partial_s\theta+y(1-M\partial_s\theta)+M\cos\theta=0
\label{halffinal}
\end{equation}
This equation can be solved numerically to obtain the profile of the film.

To compare \eqref{halffinal} with the final equation of Diamant and Witten \cite{Diamant2011SI}, we differentiate with respect to $s$:
\begin{equation}
\partial_s^4\theta + \partial_s^2\theta \left[\tfrac{3}{2}\left(\partial_s\theta\right)^2+P-My\right]+\sin\theta(1-2M\partial_s\theta)=0
\label{final}
\end{equation}
When $M=0$, equation \eqref{final} reduces to that derived by Diamant and Witten \cite{Diamant2011SI}. \\

\paragraph*{Boundary conditions}
To model our experiments, the appropriate boundary conditions are $y(\pm L_0/2)=0$. However, for an idealized, infinite film the boundary condition $y(\pm \infty)=-M$ might be more appropriate (so that it is freely floating far from the localization). This modification would change one term in the final equation, however this term can be absorbed through a shift in the load \cite{Oshri2015SI} to recover equation \eqref{final}.

\section{Fold symmetries}
\begin{figure}[htbp]
\centering
  \includegraphics[width=0.5\textwidth]{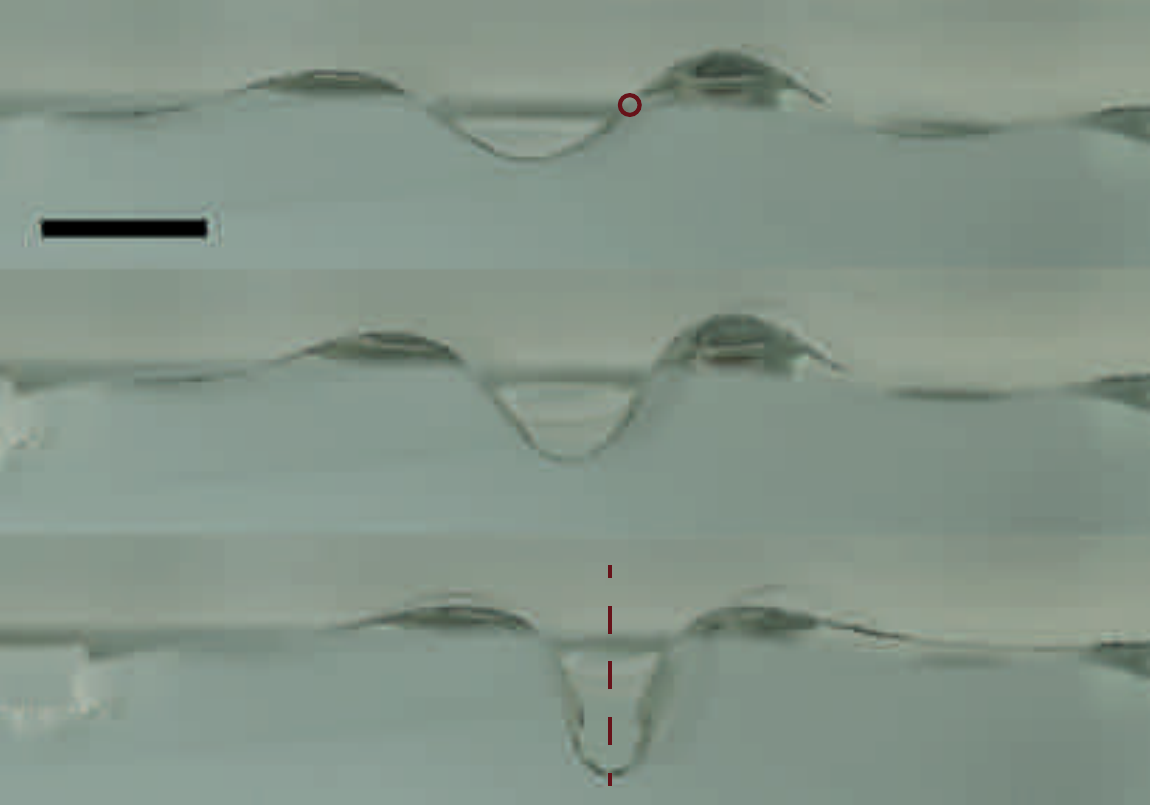}
  \caption{Side pictures of an elastic film at the oil/water interface, illustrating the buckling route ($\rho_s=1.4\mathrm{~g\,cm^{-3}}$, $M=0.10$, shown by the orange curves in Fig.~3 of the main text). Top picture: antisymmetric (centre of rotation highlighted), middle picture: no symmetry and bottom picture: symmetric (reflection axis drawn). Compression increases from top to bottom with $\Delta/\lambda=0.20,\:0.33,\:0.49$, respectively. Scale bar: 5mm.}
  \label{fgr:sym}
\end{figure}

Pocivavsek \textit{et al} \cite{Pocivavsek2008SI} identify four possible buckling modes (two symmetric and two antisymmetric ones). Experimentally they observe that at high compression all their films become symmetric (mainly downward) but give no explanation for this phenomenon. Theoretical works on weightless infinite films\citep{Diamant2013SI, Rivetti2013SI} show that there are an infinity of continuous buckling modes from symmetric to antisymmetric which all have the same energy. Rivetti and Neukirch \cite{Rivetti2014SI} study finite length films numerically and show that, depending on the film length, there is one stable buckling mode for a given compression. In a numerical experiment in which the compression is increased gradually, the film can transition from antisymmetric to symmetric (and vice versa) configurations several times via non symmetric modes. They call this the branching route. In our experiments we observe such transitions (Fig.~\ref{fgr:sym}). Two possible branching routes are observed: the wrinkles start symmetric and the fold remains symmetric until self-contact or the wrinkles start antisymmetric (or non symmetric) and the film goes through a series of non symmetric modes until it reaches a downward symmetric fold and then stays downward symmetric until self-contact. We never observe more than one transition, even though multiple transitions have been predicted theoretically. At high compression, the film weight (neglected in previous theoretical work) stabilizes the downward symmetric configuration, making it energetically favourable over other modes.

Finally, the initial state of the film is very sensitive to the boundary conditions: if the film is not perfectly aligned, the symmetry of the profile and the fold position are noticeably modified. Moreover, small defects (in the thickness or material properties) can modify the buckling modes making  prediction of the precise buckling route in a practical application difficult.

\section{Role of surface tension}

Surface tension is neglected in most works on floating thin films since this assumption allows the problem to be simplified to a two dimensional analysis. Nevertheless, small capillary bridges exist at the edges of the film which exert strong forces on it. For example, it has been shown\cite{Huang2010SI} that for ultrathin films with very low bending modulus, surface tension modifies the wrinkle wavelength over a distance $\ell_c=(\gamma/\Delta\rho g)^{1/2}$. For the effect of surface tension to be negligible, therefore, the film needs to have a high bending modulus (such that $\ell_{eh}\gg\ell_c$). In the experiments presented in this paper surface tension effects are small but visible: we find a slight reduction of the wavelength (about 15\%) at the edges, with the film profiles distorted in the neighbourhood of the edge; finally, self-contact is reached sooner (i.e.~at smaller compressions). 

\begin{figure}[htbp]
\centering
  \includegraphics[width=0.8\textwidth]{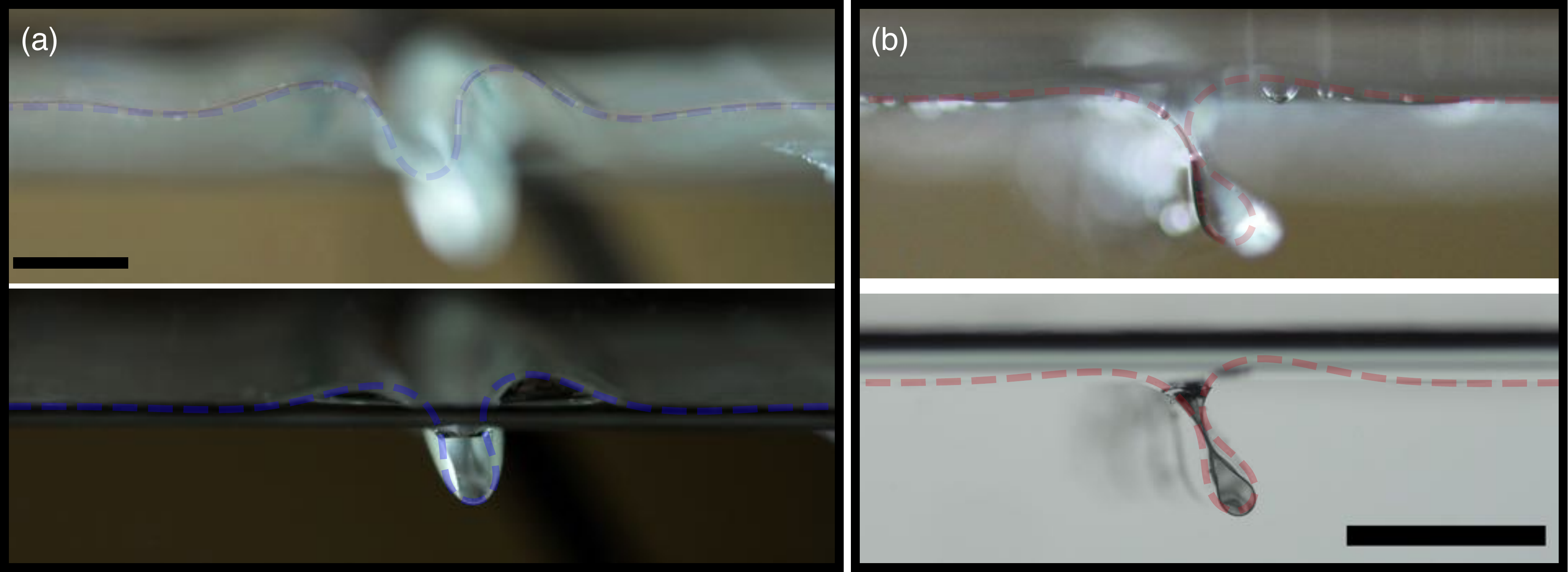}
  \caption{\textbf{(a)} Side picture of a VPS film compressed at the air/water interface viewed in the centre of the film (top) and at the edge of the film (bottom), both with the same compression. Top picture: the camera is focused on the centre of the film. The red line is physically drawn on the film, while the overlaid blue dashed curve is the theoretical solution with the experimentally measured parameters ($\Delta$, $\lambda$) and $k\phi=0.60$. Bottom picture: the camera is focus on the edge of the film, the overlaid blue dashed line is the theoretical solution with the experimentally measured parameters with $k\phi=0.35$.  \textbf{(b)} Side images of a PDMS film compressed at the air/water interface viewed at the point of self-contact (though not the same value of $\Delta$ in each case); here images are shown with the camera focussed $\sim 8\mathrm{~mm}$ from the edge and at the edge. Top picture: focus is on the centre of the film. The black line is physically drawn on the film while the overlaid red dashed curve is the theoretical solution with the experimental parameters and $k\phi=0.45$. Bottom picture: focus is at the edge, the overlaid red dashed line is the theoretical solution with the experimental parameters and $k\phi=0.45$. Scale bars: 5 mm.}
  \label{fgr:superpos}
\end{figure}

To highlight the effect of surface tension in this region, we take relatively transparent films (thin pure VPS films and polydimethylsiloxane or PDMS films) with negligible weight ($M\sim0$) and draw lines with a marker pen away from the edges. Since the films are transparent, we can focus the camera on the drawn line. Fig.~\ref{fgr:superpos} (a) shows that the profile far from the edge, and hence undisturbed by surface tension, is well fitted by the usual solution \citep{Diamant2013, Rivetti2013}. However, the profile at the edge of the film, where surface tension creates strong deformations, cannot be fitted by such solutions. Fig.~\ref{fgr:superpos} (b) shows that the teardrop shape at self-contact is accurately predicted with the solution from \citep{Diamant2013, Rivetti2013} only far from the film edges.

\section{Heavy hanging column model after self-contact}
A vertical film of length $L^*$, thickness $t$, width $W$ and density $\rho_s$ is immersed in water $\rho_w$. The film is clamped at the top and a vertical force $F$ is pushing at the bottom along the full width $W$.
We introduce the intrinsic coordinates ($s$, $\theta$) where $s$ is the arclength going from bottom to top and $\theta (s)$ is the local angle between the tangent and the vertical axis $x$ (schematic Fig.~6). The $(x, y)$ coordinates are therefore rotated clockwise by $90$ degrees compared to those used for prior to contact. We parametrize the film centreline as $[x(s),y(s)]$ which are given in terms of the intrinsic coordinates by ($\partial_sx=-\cos\theta$, $\partial_sy=-\sin\theta$). The problem is invariant along the width of the film so we can eliminate $W$ and consider the problem in two dimensions. The internal forces, $n_x(s)$ and $n_y(s)$ and moment equilibrium, $\overrightarrow{m(s)}=m(s)\vec{e_z}$ per unit of width on a small portion of the beam read as:
\begin{eqnarray}
&\partial_sn_x=-f_{x} \nonumber\\
&\partial_sn_y=-f_{y} \nonumber\\
&\partial_sm=n_y\cos\theta-n_x\sin\theta \nonumber
\end{eqnarray}
with $f$ the external force (per unit width) distributed along the beam. Here the only external force is gravity (with buoyancy taken into account), thus $f_y=0$ and $f_x=(\rho_s-\rho_w)gt$. We therefore integrate between $0$ and $s$ the first two equations:
\begin{eqnarray}
&n_x(s)-n_x(0)=-(\rho_s-\rho_w)gts \nonumber\\
&n_y(s)-n_y(0)=0 \nonumber
\end{eqnarray}
At $s=0$ there is an upward vertical force $F$ (per unit width) so that $n_y(0)=0$ and $n_x(0)=F$. So finally:
\begin{eqnarray}
&n_x(s)=F-(\rho_s-\rho_w)gts \nonumber\\
&n_y(s)=0 \nonumber
\end{eqnarray}
We insert these results into the internal moment equation and consider the film slender enough  that its local moment is proportional to the local curvature $m(s)=B\partial_s\theta(s)$, with $B$ the bending modulus of the beam (equivalent to that of the elastic film measured experimentally, which makes up a single side of the fold). We obtain the equation describing the shape of the beam:
\[\partial_s m=B \partial_s^2 \theta = -[F-(\rho_s-\rho_{w})gts] \sin \theta\]
The boundary conditions $\partial_s \theta(s=0)=0$, $\theta(s=L^*)=0$ are imposed. The system is made dimensionless by dividing $s$ by $L^*$:
 \begin{equation}
\begin{gathered}
\partial_s^2 \theta + \left[\widetilde{F}-\bigg(\frac{L^*}{\ell_g}\bigg)^3 s\right]\sin \theta =0 \\
\partial_s \theta(s=0)=0, \quad \theta(s=1)=0
\end{gathered}
\label{heavyeleq}
\end{equation}
$\widetilde{F}=\frac{F{L^*}^2}{B}$ is the dimensionless force and $\ell_g=\left(\frac{B}{(\rho_s-\rho_{w})gt}\right)^{1/3}$ is the elasto-gravitational length which is the characteristic length at which the weight of the film is enough to cause it to bend.

To find the onset of buckling we consider small deformations and linearize equation \eqref{heavyeleq} \textit{i.e.} we let $\sin\theta \approx \theta$ giving
\[\partial_s^2 \theta + \left[\widetilde{F}-\bigg(\frac{L^*}{\ell_g}\bigg)^3 s\right]\theta =0\]
We introduce the new arclength variable $r=\frac{L^*}{\ell_g}\bigg(s-\widetilde{F}\Big(\frac{\ell_g}{L^*}\Big)^3\bigg)$, so that the equation becomes:
\begin{equation}
\partial_r^2\theta-r\theta=0,
\label{heavyeleqlin}
\end{equation}
which has general solution
\begin{equation}
\theta(r)=C_1  \operatorname{Ai}(r)+C_2 \operatorname{Bi}(r)
\end{equation} with $\operatorname{Ai}$ and $\operatorname{Bi}$ the Airy functions of the first and second kind and the constants $C_1$ and $C_2$ are determined by the boundary conditions, which in the rescaled coordinates used here read
\begin{equation}
\partial_r \theta \bigg(r=-\widetilde{F}\Big(\frac{\ell_g}{L^*}\Big)^{2/3}\bigg)=0, \quad \theta \bigg(r=\frac{L^*}{\ell_g}\Big(1-\widetilde{F}\left(\frac{\ell_g}{L^*}\right)^3\Big)\bigg)=0. \nonumber
\end{equation}
 For non trivial solutions to exist ($\theta(s)\neq0$, corresponding to buckling of the column) we need:
\begin{equation}
\begin{vmatrix}
\operatorname{Ai}\Big[\frac{L^*}{\ell_g}\Big(1-\widetilde{F}\Big(\frac{\ell_g}{L^*}\Big)^3\Big)\Big]& \operatorname{Bi}\Big[\frac{L^*}{\ell_g}\Big(1-\widetilde{F}\Big(\frac{\ell_g}{L^*}\Big)^3\Big)\Big]\\
\operatorname{Ai}'\Big[-\widetilde{F}\Big(\frac{\ell_g}{L^*}\Big)^2\Big]& \operatorname{Bi}'\Big[-\widetilde{F}\Big(\frac{\ell_g}{L^*}\Big)^2\Big]
\end{vmatrix}=0 \nonumber
\label{heavyelthreshold}
\end{equation}
This equation is solved using \texttt{Mathematica}. The first root gives us the critical dimensionless force $\widetilde{F}_c$ to buckle (in mode 1) as a function of the dimensionless beam length $(L^*/\ell_g)$: $\widetilde{F}_c=\frac{F_c{L^*}^2}{B}=\operatorname{f}\big((\frac{L^*}{\ell_g})^3\big)$ where the function $\operatorname{f}$ is found numerically.

In our experiment, it is the beam length $L^*$ that is varied  (the buoyancy force from the teardrop is constant). Since $L^*$ appears in the  non-dimensionalization of the force, it is more convenient to write:
\begin{equation}
\frac{F\ell_g^2}{B}=\Big(\frac{\ell_g}{L_c}\Big)^2\operatorname{f}\bigg(\frac{L_c}{\ell_g}\bigg)
\label{heavyelthreshold2}
\end{equation}

The predicted critical force, based on equation \eqref{heavyelthreshold2}, is shown as the red curve in Fig.~6 of the main text. This qualitatively captures the fold's behaviour: there is a critical buoyancy force below which the fold never bends back towards the surface, regardless of the length $L^*$. However, the data points consistently lie well below the red curve described by \eqref{heavyelthreshold2}. Using equation ($5$) of the main text, the lowest critical force ($F \ell_g^2 / B \approx 1$) reduces to $M^{-2/3} (\tfrac{{\mathscr{A}}}{\ell_{eh}^2}-M\tfrac{{\mathscr{L}}}{\ell_{eh}})\approx 1$. If we use images from the side to evaluate ${\mathscr{A}}/\ell_{eh}^2$ and ${\mathscr{L}}/\ell_{eh}$, we find that the critical dimensionless mass $M=0.060$, which is below the experimentally determined value ($M\sim 0.14$).
 In most experiments the fold is not perfectly aligned with the vertical axis when it reaches self-contact (see, for example, Fig.~4(a) of the main text). The average angle $\alpha$ between the fold and the vertical axis is $\left|\alpha\right|=10^{\circ}$ (its maximum value is $26^{\circ}$).

To account for the effect of the slope of the fold at self-contact, we add an initial angle to the heavy hanging column model by changing the boundary condition at the top of the beam. We then solve equation \eqref{heavyeleq} numerically with the new boundary condition $\theta(s=1)=\alpha$ to get the beam shape. With an initial angle the transition from straight to a buckled configuration becomes smooth. We therefore need to define a criterion to describe the critical length $L^*_c/\ell_g$ for buckling. Here, we use the normalized free end horizontal displacement, $y/L^*(s=0)$ (denoted as $y_0$), to give this criterion: when the beam is straight $y_0=\sin(\alpha)$ but when the beam starts to bend $y_0$ increases (Fig.~\ref{fgr:heavyelexpdiagphase}(a)). We define the critical length $L^*_c$ as  the length at which $y_0-\sin(\alpha)>0.1$. (This choice of threshold comes from experiments: the chosen threshold has to be much higher than the measurement uncertainty so that the ``buckled'' beam can be visually identified.)

Finally, we note that the force used in Fig.~6 of the main text is, in fact, an under-estimate of the true force (since  the shape of the teardrop is measured from the side where surface tension plays an important role). We have shown in Fig.~\ref{fgr:superpos} (b) that away from the side the shape of the fold up to self-contact can be fitted by the solution in ref \citep{Diamant2013, Rivetti2013}. The width averaged force due to the encapsulated fluid is thus better estimated with the solution from ref \citep{Diamant2013, Rivetti2013} at self-contact. Fig.~\ref{fgr:heavyelexpdiagphase}(b) therefore shows the phase diagram for the fold bending using the calculated values of the teardrop shape at self-contact, instead of that measured from side images. We find a good quantitative agreement with the tilted heavy hanging column model despite the numerous approximations. Moreover, the critical dimensionless mass is now $M=0.094$, closer to the experimental value ($M\sim 0.14$).

\begin{figure*}[htbp]
\centering
  \includegraphics[width=0.95\textwidth]{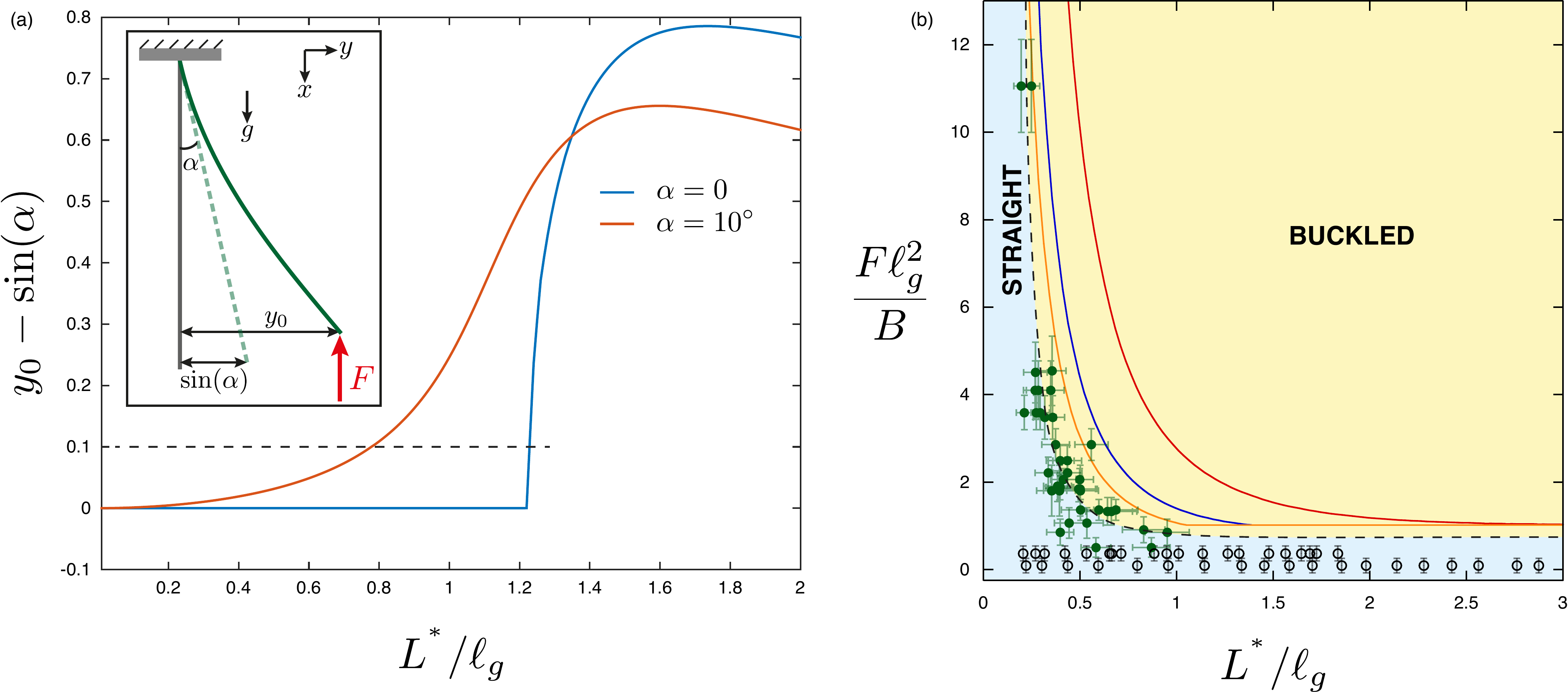}
  \caption{\textbf{(a)} Dimensionless horizontal end displacement of the beam as a function of its length for a dimensionless force $F\ell_g^2/B=2$ with two different clamping angles at the end, $\alpha$: $\alpha=10^{\circ}$ (orange curve) and $\alpha=0^{\circ}$ (blue curve). The horizontal dashed line shows our criterion for determining the critical buckling length $L^*_c$. \textit{Inset} Schematic of the tilted heavy hanging column defining the horizontal end displacement $y_0$, with all lengths normalized by $L^*$. \textbf{(b)} Phase diagram of the post self-contact fold buckling using the solution of Diamant and Witten\cite{Diamant2011SI} to compute the teardrop area and perimeter. Green filled circles give the critical length $L^*$ before buckling for each experiment. Black open circles represent experiments in which the fold never buckled. All parameters are varied in the data presented. The black dashed curve and background colours are guides for the eye to distinguish experimental phases. The red solid curve is the analytical result of the heavy hanging column theory \eqref{heavyelthreshold2}, the blue (respectively orange) solid curve are the numerical results of the heavy hanging column theory with an initial angle of $10^{\circ}$ (respectively $20^{\circ}$). }
  \label{fgr:heavyelexpdiagphase}
\end{figure*}

\section{Film characterization}
\setcounter{figure}{1} 
Table \ref{tab:sheetprop} shows the  densities and thicknesses of the films used in our experiments.

\begin{table}[!tb]
	\centering
	\begin{tabular}{ccc r@{ $\pm$ }l r@{ $\pm$ }l r@{ $\pm$ }l  r@{ $\pm$ }l}
		\toprule
		$\rho_s$ & $W$ & $L$ & \multicolumn{2}{c}{$t_{weight}$} &\multicolumn{2}{c}{$t_{laser}$} & \multicolumn{2}{c}{$M_{air/water}$} &\multicolumn{2}{c}{$M_{oil/water}$}\\
		($\mathrm{g\,cm^{-3}}$) & ($\mathrm{mm}$) & ($\mathrm{mm}$) & \multicolumn{2}{c}{($\mathrm{\mu m}$)} & \multicolumn{2}{c}{($\mathrm{\mu m}$)} & \multicolumn{2}{c}{} & \multicolumn{2}{c}{} \\ \hline \midrule
		1.20 & 50 & 75 & 74 & 6 & 54 & 4 & 0.007 & 0.001 & 0.030 & 0.004 \\
		1.40 & 50 & 75 & 81 & 5 & 90 & 11 & 0.023 & 0.004 & 0.100 & 0.019 \\
		1.40 & 60 & 90 & 89 & 4 & 89 & 14 & 0.023 & 0.005 & 0.091 & 0.020 \\
		1.54 & 50 & 75 & 73 & 4 & 79 & 11 & 0.026 & 0.005 & 0.104 & 0.021 \\
		1.54 & 60 & 90 & 89 & 4 & 109 & 12 & 0.031 & 0.005 & 0.131 & 0.022 \\
		1.78 & 50 & 75 & 79 & 4 & 78 & 6 & 0.039 & 0.005 & 0.144 & 0.020 \\
		1.78 & 60 & 90 & 90 & 4 & 98 & 9 & 0.045 & 0.007 & 0.178 & 0.026 \\
		2.00 & 50 & 75 & 85 & 4 & 92 & 10 & 0.051 & 0.008 & \multicolumn{2}{c}{} \\
		2.00 & 60 & 90 & 91 & 3 & 104 & 7 & 0.054 & 0.007  & \multicolumn{2}{c}{} \\
		2.12 & 50 & 75 & 83 & 3 & 99 & 12 & 0.058 & 0.01 & \multicolumn{2}{c}{} \\
		2.12 & 60 & 90 & 94 & 3 & 117 & 28 & 0.061 & 0.018 & \multicolumn{2}{c}{} \\
		2.36 & 50 & 75 & 64 & 3 & 80 & 8 & 0.059 & 0.009 & \multicolumn{2}{c}{} \\
		2.36 & 60 & 90 & 63 & 2 & 70 & 6 & 0.060 & 0.009 & \multicolumn{2}{c}{} \\
		2.55 & 50 & 75 & 49 & 2 & 64 & 7 & 0.078 & 0.013 & \multicolumn{2}{c}{} \\
		2.55 & 60 & 90 & 55 & 2 & 69 & 5 & 0.077 & 0.010 & \multicolumn{2}{c}{} \\ \bottomrule 
	\end{tabular}
	\caption{Film properties. The values of $M$ displayed are calculated using $M=2\pi(\rho_s-\rho_w)t_{laser}/(\Delta\rho \lambda)$.}
	\label{tab:sheetprop}
\end{table}

We use a tensile test to measure the Young's moduli and Poisson ratios of the materials used. In particular, the Young's modulus is extracted from a linear fit of the true stress as a function of the longitudinal strain, while the Poisson ratio is measured from the lateral strain as a function of the longitudinal strain. We find for the pure VPS dogbone that $\nu=0.46$ and $E=1.03\mathrm{~MPa}$ while for a dogbone of VPS with iron powder $\nu=0.49$ and $E=2.88\mathrm{~MPa}$. The difference in Poisson ratio is not significant and we take the value $\nu=0.5$ for all our films. However, the Young's modulus increases significantly as  iron powder is added.

We also check that the values of the Young's modulus from the tensile tests are consistent with a beam deflection test on two films: one made of pure VPS $\rho_s=1.20\mathrm{~g\,cm^{-3}}$, the other one made of VPS mixed with iron powder $\rho_s=2.39\mathrm{~g\,cm^{-3}}$. We clamp the film horizontally and let a length $L$ hang freely under its own weight. In this configuration the deflection at the end of the sheet is given by $y_{end} =\rho_s t L^4 g/(8 B) = 3 (1-\nu^2) \rho_s g L^4/(2 E t^2) $. We extract $E$ from a linear fit of $y_{end}$ as a function of $L^4$, obtaining $E=1.46 \pm 0.24 \mathrm{~MPa}$ for the pure VPS film and  $E=2.56 \pm 0.47\mathrm{~MPa}$ for the film made of VPS with iron powder.

\end{document}